\documentclass[aps,twocolumn,amsmath,amssymb,floatfix,superscriptaddress,nofootinbib]{revtex4-1}

\usepackage{graphicx}
\usepackage[caption=false]{subfig}
\usepackage{dcolumn}
\usepackage{bm}
\usepackage{latexsym}
\usepackage{color}
\usepackage[colorlinks=true,linkcolor=blue,citecolor=blue]{hyperref}
\usepackage{xcolor}
\usepackage{lipsum}

\def\mnras{{Mon.\ Not.\ Royal\ Astron.\ Soc.}}

\begin{document}

\title{Particle Acceleration in Colliding Flows: Binary Star Winds and Other Double-Shock Structures }

\author{Mikhail Malkov}
\email{mmalkov@ucsd.edu}
\affiliation{Department of Physics and CASS, University of California San Diego,
La Jolla, CA 92093, USA}
\author{Martin Lemoine}
\affiliation{Institut d'Astrophysique de Paris, CNRS -- Sorbonne
Universit\'e, F-75014 Paris, France}

\date{\today}

\begin{abstract}  
A shock wave propagating perpendicularly to an ambient magnetic field
accelerates particles considerably faster than in the parallel propagation
regime. However, the perpendicular acceleration stops after the shock
overruns a circular particle orbit. At the same time, it may continue
in flows resulting from supersonically colliding plasmas 
bound by a pair of perpendicular shocks. Although the double-shock
acceleration mechanism, which we consider in detail, is not advantageous
for thermal particles, pre-energized particles may avoid the premature
end of acceleration. We argue that if their gyroradius exceeds the
dominant turbulence scale between the shocks, these particles might
traverse the intershock space repeatedly before being carried away
by the shocked plasma. Moreover, entering the space between the shocks
of similar velocities $u_{1}\approx u_{2}\approx c$, such particles
start bouncing between the shocks at a fixed angle $\approx\negthinspace\negthinspace35.3^{\circ}$
to the shock surface. Their drift along the shock fronts is slow,
$V_{d}\sim\left|u_{2}-u_{1}\right|\ll c$, so that it will take $N\sim Lc/\left|u_{2}-u_{1}\right|d\gg1$
bounces before they escape the accelerator (here, $L$ is the size
of the shocks and $d$ is the gap between them). Since these particles
more than ten-fold their energy per cycle (two consecutive bounces),
we invoke other possible losses that can limit the acceleration. They
include drifts due to rippled shocks, the nonparallel mutual orientation
of the upstream magnetic fields, and radiative losses. 
\end{abstract}

\maketitle

\section{Introduction}

Colliding supersonic flows are ubiquitous to astrophysical phenomena. Typically, they form a layer of shocked plasma bound by a pair of termination shocks propagating back into the respective flows. 
Such a flow configuration may occur explosively, as in supernova remnants (SNRs)~\cite{Reynolds_2008}, where they find one of their most striking manifestations. There, 
a pair of forward-reverse shocks and a radiatively cooled outer shell are formed in the early and late evolution phases, respectively.  A double-shock configuration may also occur under continuous energy release, for instance, through the collision of stellar outflows in binary systems \cite{Usov1992,1993ApJ...402..271E,2006MNRAS.372..801P}, whether composed of massive stars or of massive stars and a compact companion, see, {\it e.g.,} ~\cite{2013A&A...558A..28D,2013A&ARv..21...64D,2019frap.confE..44P} for reviews. 

Isolated shocks have attracted much attention as efficient particle accelerators, e.g.~\cite{87Blandford} and references  therein, therefrom, but less so did the double-shock systems in colliding flows~\cite{2013MNRAS.429.2755B,2021ApJ...921L..10B}. Recent detections at very high energies have, however, triggered renewed interest; see, for instance, the massive star binary $\eta$~Car~\cite{2009ApJ...698L.142T} or the gamma-ray binary LS~5039~\cite{2006A&A...460..743A}.
There are, a priori, physical obstacles for them to accelerate particles more efficiently than isolated
shocks. First, a growing amount of shocked material increases
the distance between termination shocks to a point beyond which they
accelerate particles quasi-independently. However, this distance does not
always increase significantly. For example, in colliding winds of
binary stars, the lateral outflow of the shocked plasma keeps the
shock separation constant; for numerical simulations, see~\cite{2011MNRAS.418.2618L,2015A&A...581A..27D,2008MNRAS.387...63B,Bogovalov2012,2012A&A...544A..59B,2018MNRAS.479.1320B,2019MNRAS.490.3601B,2020A&A...641A..84M}. But sideways
downstream outflows also create problems for particle acceleration
in the double-shock regime. 

The major problem is that particles couple to the downstream flow
by scattering on magnetic perturbations, which is actually vital for
the single-shock acceleration. In double-shock systems, the shocked
plasma typically diverts from a stagnation point, thus moving largely
along the shock fronts. Accelerated particles will then be convected
away from the acceleration region if they couple to the flow. However,
for the acceleration to be efficient within the diffusive (single)
shock acceleration, particles must drive the perturbations by themselves,
almost automatically matching the perturbation scales to the particle
energies (gyro-radii). Hence, a limited number of high-energy test
particles, not capable of driving their own waves, will not be captured
by the local fluid element. Perturbations excited by the bulk of accelerated
particles will be shorter than the giroradii of these test particles and will produce
only a shiver effect on their orbits. Such particles are not convected
toward the edge of shocked flow or even significantly deflected from
almost rectilinear trajectories between the shocks. If, in addition,
the shocks are quasi-perpendicular, a particle will not be able to escape
upstream farther than its gyroradius. Hence, it will effectively bounce
off the upstream field on either side of the shocked layer, thus not
driving waves upstream in the first place. Under a ``standard''
diffusive shock acceleration (DSA) paradigm, these waves provide the
scattering environment for the downstream particles after being convected
with the flow across the shock interface. 

The above-described particle kinematics radically shortens the acceleration
cycle compared to the standard DSA. The acceleration mechanism is
thus similar to the classical Fermi acceleration of a particle bouncing
between two plates that approach each other but never collide. 
This configuration advantageously combines the short acceleration
cycle of perpendicular shocks with the long confinement time of parallel
ones.

Another example where the thickness of the shocked layer can be limited
is a radiative shell of an old SNR, {\it e.g.}~\cite{Chevalier1974}.
Here the shocked layer does not thicken significantly because the
internal energy of the shocked plasma is promptly radiated away so
that the bounding shocks overcompress the gas layer between them.
This happens because of the neutral gas that is quickly cooled off
by the line emission. The shocked layer stability against shock rippling,
bending as a whole, \textquotedblleft bloating\textquotedblright,
and Kelvin-Helmholtz shear flow instability has been studied both
in astrophysical \cite{Vishniac1983ApJ,Vishniac1994ApJ,ChevalierBWinst1992ApJ,Stevens1992,BlondinNTSI1996}
and laboratory \cite{Sarkisov2005} contexts. We will focus on the
shock rippling that strongly affects particle acceleration and their
propagation along the shocked layer. These effects, along with the
noncollinearity of the magnetic fields upstream, will be key in terminating
particle acceleration.

Further interest in double-shock systems is driven by a limited range
of particle spectral indices predicted by the DSA. This constraint often
hampers modeling. In nonrelativistic shocks, the particle momentum
spectrum (pitch-angle averaged and normalized to $p^{2}dp$), scales
as $\propto p^{-q}$. The index $q$ is robustly determined by the
shock compression $r$, $q=3r/\left(r-1\right)\ge4$. Even if the shock is modified by the pressure of accelerated
particles, leading to the shock compression well above four, the index
remains limited by $q_{\text{min}}=3.5$ \cite{m99}, not three,
as predicted by the above formula for $q$. In relativistic shocks,
the spectrum has long been regarded being even softer, $q\approx4.2$~\cite{1998PhRvL..80.3911B,KirkDuffyRelS99,GallantAcht99,Achterberg2001,2003ApJ...589L..73L}.
Additionally, there is mounting (albeit, indirect) evidence from multi-messenger studies for the need of hard injection spectra. The so-called ``extreme'' blazars, for instance, reveal photon spectra at TeV with indices harder than $\sim2$, meaning $F_\nu/\nu \propto \nu^{-\Gamma}$ with $\Gamma<2$~\cite{2020NatAs...4..124B}. If taken at face value, this suggests an index $q<4$ for the electron spectrum, although various radiative effects may affect the relationship between $q$ and $\Gamma$. On yet another front, the modeling of the cosmic-ray spectra at the highest energies seemingly requires hard spectra ($q<4$) as well, see {\it e.g.} \cite{2017JCAP...04..038A}. Here as well, of course, the connection between the propagated spectrum and that at injection can be plagued by various energy-dependent effects such as cosmic-ray escape.

Being synergistic in nature, the double-shock acceleration still relies on 
the particle interaction with individual shocks, which has formed the
focus of numerous research papers and reviews, particularly devoted
to ultrarelativistic shocks \cite{KirkDuffyRelS99}. 
Previous work by A. Bykov and collaborators~\cite{2013MNRAS.429.2755B,2021ApJ...921L..10B} have demonstrated that hard spectra could arise naturally at double-shock systems of colliding wind binaries, with potentially important phenomenological consequences for the generation of high-energy ($\sim\,$PeV) neutrinos and photons from Galactic sources. The acceleration becomes particularly efficient in the energy range in which particles can probe both shock systems, as in the present study. We concentrate here on the kinematics of the energization process, focusing on
the energy
gain of particles hitting the shock from behind and deflected back
downstream. As we mentioned, there are both relativistic and nonrelativistic
double shock systems, but also the mixed ones~\cite{2011MNRAS.418.2618L,2015A&A...581A..27D,2008MNRAS.387...63B,Bogovalov2012,2012A&A...544A..59B,2018MNRAS.479.1320B,2019MNRAS.490.3601B,2020A&A...641A..84M}.
Therefore, we will address the shock-particle interaction, not constrained
by the shock speed.
\begin{figure*}
\includegraphics[viewport=80bp 400bp 612bp 765bp,scale=0.54]{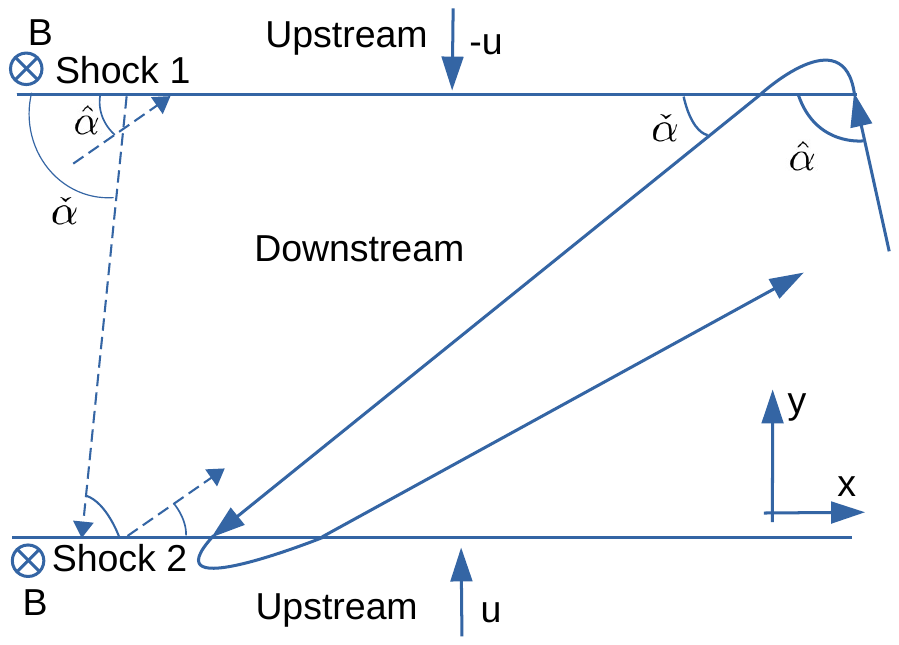}\includegraphics[viewport=100bp 310bp 612bp 765bp,scale=0.54]{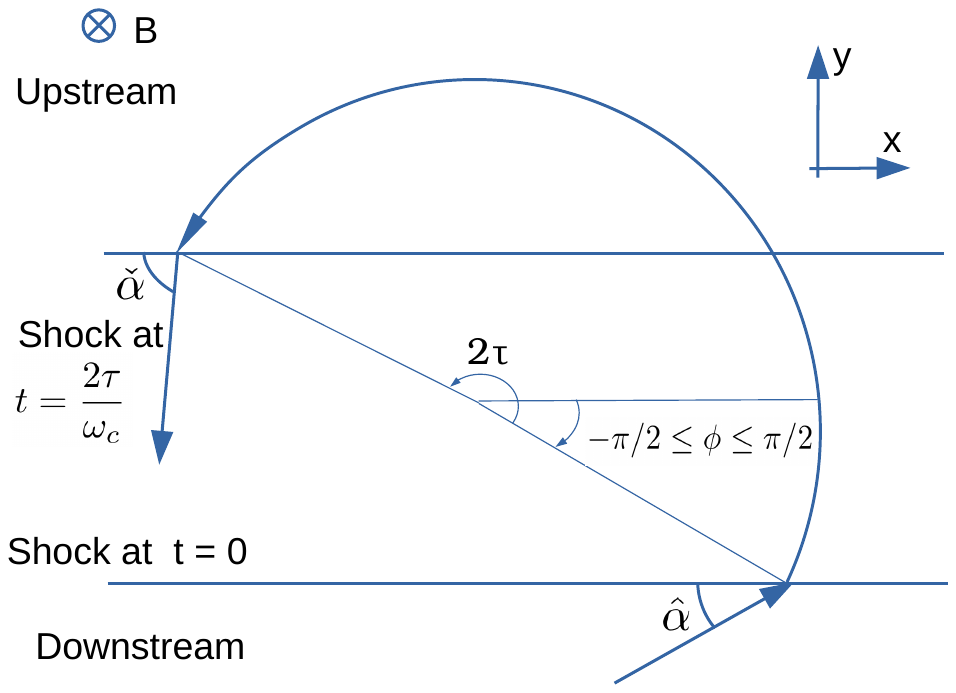}
\caption{\textbf{Left panel}: Schematics of particle bounces between two shocks.
The magnetic field is perpendicular to the figure plane upstream of both
shocks. \emph{Solid line}: a general-case orbit, entering the shock-1
from downstream at angle $\hat{\alpha}$, turning around, exiting
at $\check{\alpha},$ entering shock-2 at the same angle $\check{\alpha}$,
etc. \emph{Dashed line}: a possible periodic orbit in which a particle
alternates the same pair of angles $\hat{\alpha}\protect\neq\check{\alpha}$,
thus drifting along the shocks because the angles are different.\textbf{
Right panel}: Particle orbit upstream between front crossing and recrossing.
It enters upstream at angle $\hat{\alpha}$ in the shock frame, rotates
by $2\tau$ between angle $\phi$ and $\phi+2\tau$ in the upstream
frame and recrosses the shock in the downstream direction (see text).  This figure is not to scale: the distance between the two shocks is assumed to be smaller than the gyroradius of the particles in the inflowing regions.
\label{fig:Particles-bouncing-between}}
\end{figure*}

Finally, it is of interest to remark that colliding flows also occur in reconnection layers, on scales many orders of magnitude smaller than those of stellar binaries~\footnote{We kindly acknowledge one of our referees for pointing out this connection.}.  In a symmetric Harris-type configuration, the plasma inflows from both sides of the current layer 
at a velocity $v_{\rm rec}\simeq 0.1\,v_{\rm A}$, which can take values as large as $\sim 0.1\,c$ at large magnetizations, i.e. when the Alfv\'en velocity $v_{\rm A}\sim c$~\cite{2015SSRv..191..545K}. 
In that relativistic regime, recent 3D simulations have demonstrated that high-energy particles 
gain energy precisely by bouncing across the layer~\cite{2021ApJ...922..261Z}. 
Here, ``high-energy'' means Lorentz factors above $\sigma \equiv v_{\rm A}/\sqrt{1-v_{\rm A}^2/c^2}$, at which point the gyroradius of the particles exceeds the thickness of the current sheet. This reconnection geometry thus offers a vivid illustration of the setup we have in mind in the present paper, namely a particle traveling in a nearly rectilinear fashion between two colliding flows carrying perpendicular magnetic fields; see~\cite{2010MNRAS.408L..46G,2012A&A...542A.125B} for discussions of particle acceleration in that particular context. An important parameter that we discuss more specifically in Section~V.C is the angle $\theta\in[0,\pi]$ between the magnetic field directions. In a reconnection setting, it is directly related to the relative magnitudes of the guide field $B_0$ 
and the reconnecting component $B_{\rm rec}$, via $\cos\theta = \left(B_0^2-B_{\rm rec}^2\right)/\left(B_0^2+B_{\rm rec}^2\right)$. 

\section{Framework of Double-Shock Acceleration}\label{sec:Double-Shock-Acceleration-Model}

We approach the particle acceleration in a double-shock system by
constructing an \emph{iterated map}. It starts from the particle relativistic
factor $\gamma$ and its two incidence angles relative to the shock
plane and magnetic field direction before hitting one of the shocks,
as shown in Fig.\ref{fig:Particles-bouncing-between}. Let the projection
of the ingress angle to the plane transversal to the magnetic field
be $\hat{\alpha}$. The particle pitch angle to the magnetic field,
$\cos^{-1}\mu$, is not shown because the field is perpendicular to
the plane of the figure. The angle $0<\hat{\alpha}<\pi$ is counted
counterclockwise from the shock surface on the plane perpendicular
to the magnetic field. We assume that no significant scattering occurs
to high-energy test particles upstream, as the shock is perpendicular
and more abundant particles of lower energies do not penetrate far
enough upstream to disturb the inflowing plasma at scales significant
for the high-energy test particles we are interested in. So, the particle
trajectory in the upstream frame is circular, and we follow it until
it recrosses the shock back in the downstream direction. The particle
egress angle downstream is denoted by $\check{\alpha}$. 

For most of our discussion,
we simplify the problem by assuming that the particle travels in a
rectilinear fashion in between the shocks but gyrates in the background
fields of the inflowing plasmas elsewhere. This corresponds to the
hierarchy $d\ll r_{\text{g}0}\ll L$, where $d$ represents the distance
between the two shocks, $r_{\text{g}0}$ is measured by the unperturbed
field, $B_{0},$ and $L$ is the distance between the two objects,
such as the distance between the stars in a binary system. This hierarchy
depends on the type of objects to be considered, as $d$ is governed
by hydrodynamical considerations, namely lateral escape of the flow,
while $r_{\text{g}0}$ is related to the energy of the particle and
to the magnetic field intensity. We also note that the lateral size
of the system, $L_{\text{sh}}$ is of the order of $L$. As particles
with $r_{\text{g}0}\gtrsim L_{\text{sh}}$ escape the system within
one gyration, the value of $L_{\text{sh}}$ sets a strict upper bound
on the gyroradius (hence energy) that can be achieved.

As regards the impact of turbulence in the inter-shock region, it can be evaluated as follows. 
When entering the intershock gap, a test particle is assumed to encounter
a plasma with the turbulent field directions correlated only on scales
much shorter than the particle's local Larmor radius, $l\ll r_{g}.$ An average
angular deflection after traversing the gap between the shocks, $d$,
will then be $\Delta\vartheta\sim\sqrt{ld}/r_{g}$,  assuming that it is accumulated from $N\sim d/l$ uncorrelated r.m.s.
deflections $\delta\vartheta^{2}\sim l^{2}/r_{g}^{2}$. Rigorous approaches
to particle transport in the short-scale turbulent field can be found
in \citep{Plotnikov_2011} and in Appendix \ref{sec:Appendix-A}. We also discuss there
a relation between the above straight-line propagation criterion $\Delta\vartheta\ll1$
and $d\ll r_{g0}$, mentioned in the preceding paragraph. Besides,
these deflections from the straight-line can be included in the iterated
map as a random variable or a regular deflection in the $B_{0}$-component.
In Sec.\ref{sec:Route-to-Chaos} we will support our choice of not including them
by identifying other deflection mechanisms.

After an ingress-egress angle transformation $\hat{\alpha}\to\check{\alpha}$
at the first shock and traversing the intershock gap, a particle executes
a half-cycle of the iterated map. The full cycle is completed by essentially
making the same steps at the second shock.  However, orbit transformations other than the
neglected deflections may be necessary when the particle strides
the gap. For example, the magnetic fields upstream of the shocks may not
be parallel, thus making a magnetic shear. Assuming
they are both parallel to the shock planes, the necessary
transformation before crossing the second shock is a rotation of the coordinate
system around the shock normal by the angle between the field directions.
Besides, the shock surfaces may be corrugated so that the shocks are
not locally parallel. We will handle such a situation
by rotating the reference frame around the respective field direction
before repeating the above-described elements of the map on the second
shock. These additional transformations can be conveniently incorporated
in a more general matrix form version of the iterated map given in
Appendix \ref{sec:Appendix-B}.

Some salient features of the iterated map method, including an onset
of stochasticity deserve a brief digression as this method is rarely
used in studies of particle acceleration in shocks; see, however \cite{MD06} or \cite{2003ApJ...589L..73L} for its application to the relativistic regime.
In our case, the iterated map is equivalent to the Poincar\'e surface
of section in phase space, familiar from the dynamical system theory.
It computes subsequent values of the particle's dynamical variables as
the orbit crosses the Poincar\'e surface. We will use one of the two
shocks or each of them as such a surface, depending on the situation.
Significant differences exist between an iterated map and continuous
integration of the full particle orbit in time. Starting from a technical
one, the map breaks down the full orbit in pieces, each of which is
sought to be computed ``exactly''. In our case, the part of particle
trajectory upstream under scatter-free conditions and a constant magnetic
field is a simple arc and can be mapped analytically. The particle
trajectory between the shocks is assumed to be rectilinear,  which we justified above.
Under these assumptions,
the map is fully deterministic at each step. Remarkably, this determinism
does not preclude a chaotic behavior of iterated map in the long run,
which we will demonstrate in Sec.\ref{sec:Route-to-Chaos}.

The onset of chaos can often be traced back to a ``memory loss''
of dynamical variables between the consecutive points of the map,
resulting from the orbit instability. These variables may then be
effectively considered random, even though they are fully deterministic.
The uncertainty occurs as the ``exact'' variables are still very
sensitive to each other and/or the system's control parameters. This
sensitivity lays the ground for orbit instability and bifurcations.
For example, a particle may be trapped into or detrapped from a wave
by an infinitesimal orbit perturbation or change of the particle energy.
It then continues along a very different trajectory after its interaction
with the wave, provided that it moves very close to a separatrix initially.
This phenomenon is often called a stochastic instability or intrinsic
instability of iterated map. 

It is remarkable that intrinsic stochasticity may develop in dynamical
systems with only one degree of freedom. This is in a sharp contrast
to the continuous dynamical systems, for which the famous Poincar\'e-Bendixson
theorem precludes chaotic dynamics unless the number of degrees of
freedom is larger than two, not even larger than one. The reason for
that is intuitively obvious and topological in nature, as a continuous
orbit on a plane cannot intersect itself, whenever the uniqueness
and smoothness of the solution is warranted by a properly constrained
``right-hand-side'' of the dynamical system. Discrete maps, on the
contrary, can go chaotic even in one dimension (e.g., the logistic
map, discussed in Sec.\ref{sec:Route-to-Chaos}), as the discrete
points may leap-frog over their earlier iterations (unlike the continuous
orbit on a plane or line). 

In a simple case of parallel magnetic fields upstream of two shocks\footnote{For the sake of clarity, let us emphasize that we assume, until Sec.\ref{subsec:Nonparallel-Fields}, that the magnetic fields are perpendicular to the shock normals; the term ``parallel'' refers here to the relative orientation between the two magnetic fields upstream of both shocks.}
our system is effectively two-dimensional, as it is sufficient to
map only $\gamma$ and $\alpha$. Moreover, it shows an asymptotically
universal, $\gamma$- independent chaotic dynamics of the particle
ingress angle $\alpha$, fully identical to one-dimensional maps.
The evolution of $\gamma$ is reduced to a monotonic increase or can
be suppressed by the particle radiation losses, if included. 

Iterated maps have significant advantages in studies of long-time
particle dynamics over the continuous-time numerical integration
since they are void of numerical discretization errors by design.
Therefore, they can evolve orbits \emph{ad infinitum}, thus allowing
one to determine the asymptotic particle dynamics. In particular,
they can definitively answer the question of whether the particle
transport in a given scattering environment is diffusive and
isolate regular (e.g., periodic) orbits from chaotic ones. The intrinsic
stochasticity results from a rapid separation of orbits close to each
other initially. It is intimately related to phase space mixing
and ergodicity. If the stochastic instability is strong, perturbations
unaccounted for in computing the particle orbit between the Poincar\'e
section crossings can be neglected compared to the intrinsic meandering
of the point on the section. At the same time ``dynamically chaotic''
orbit may be regarded as not genuinely chaotic because it is numerically
reproducible. However, adding tiny random perturbations to
the map will make the orbit irreproducible without changing the overall
parameters of the dynamically chaotic orbit (e.g., particle diffusivity,
propagation speed, etc.). Therefore, in strongly chaotic regions of
the particle phase space, powerful Ans\"atze of ergodic theory can
be applied. 

The above remarks justify the neglect of uncontrollable elements of
particle dynamics, primarily the particle scattering that occurs during
its motion between the shocks. When a particle moves on a strong chaotic
attractor of the iterated map, changes between the subsequent Poincar\'e
sections, or the lack thereof (fixed points of the map) determine
the long-time evolution. In this case, a deterministically chaotic orbit
is not significantly affected by neglected random perturbations but
becomes genuinely chaotic because of them.

\section{Iterated Map for Particle Acceleration}\label{sec:IteratedMap}

To lighten the notation, we will reuse the names of a particle's dynamical
variables when it interacts with either of the two shocks. Before
it enters the second shock, we rotate the coordinate system by angle
$\pi$ around the magnetic field direction upstream of the first shock,
in the case of plane and parallel to each other shocks. If this is
not the case and one or both shocks are corrugated, this angle will
be adjusted, assuming that the corrugation wave number is directed
across the field. If the fields upstream of the two shocks are not
parallel to each other, we also rotate the coordinate system about
the shock normal by an appropriate angle. After these steps, the field
upstream of the second shock will be perpendicular to the $x,y$ plane
and the shock will be on the top, again, as shown in Fig.\ref{fig:Particles-bouncing-between}.
We then reuse the transformations of particle momentum applied to
the first shock when the particle visits the second shock, thus completing
the full acceleration cycle. These transformations are formalized
in Appendix \ref{sec:Appendix-B}. 

Now we turn to the nomenclature of the iterated map. At each shock,
we will use two frames of reference: the rest frame of the upstream
fluid and the downstream space frame, which coincides with the rest
frames of both shocks. As we discussed in the Introduction, the shocks are separated by a fixed gap $d$. We denote the
particle momentum downstream before it enters the top shock by $\hat{\boldsymbol{p}}$.
The same momentum in the upstream frame will be $\boldsymbol{p}$.
After the particle makes an arc and is about to recross the shock
back downstream, its momentum (still in the upstream frame) becomes
$\boldsymbol{p^{\boldsymbol{\prime}}}.$ Finally, the same momentum
in the downstream frame is denoted by $\breve{\boldsymbol{p}}$. Thus,
we have the following sequence of transformations, characterizing
the particle interaction with one shock: 
\begin{equation}
\hat{\boldsymbol{p}}\to\boldsymbol{p}\to\boldsymbol{p^{\prime}}\to\breve{\boldsymbol{p}}.\label{eq:MapSeries}
\end{equation}
The particle velocity $\boldsymbol{v}$ is measured in units of $c$,
while its momentum - in $mc$. Besides $v$ and $p$, we will also
use $\gamma=\sqrt{1+p^{2}}=\sqrt{1+p_{z}^{2}+p_{x}^{2}+p_{y}^{2}},$
with $p_{z}=const$ (momentum component parallel to the magnetic field,
being constant throughout the interaction with one or both shocks
if the magnetic fields upstream of them are collinear). We extend
the above conventions introduced for $\boldsymbol{p}$ ( ``hats'',
``primes'', and ``checks'') to $\gamma$, $\boldsymbol{v}$, and
particle-shock ingress and egress angles $\alpha$, which, however,
will receive further clarification later. The first transform in eq.(\ref{eq:MapSeries})
can be written as:
\begin{equation}
p_{y}=\frac{\hat{p}_{y}+u\hat{\gamma}}{\sqrt{1-u^{2}}},\;\;\;\gamma=\frac{\hat{\gamma}+u\hat{p}_{y}}{\sqrt{1-u^{2}}},\;\;\;p_{x,z}=\hat{p}_{x,z}\label{eq:Map1}
\end{equation}
As indicated in Fig.~\ref{fig:Particles-bouncing-between}, $u$ denotes the absolute value of the relative velocity (in units of $c$) between upstream and downstream at each shock.
The next transform is defined by a rotation in the positive direction
of the particle momentum until it crosses the shock again. Particles
follow circular orbits in the upstream frame, as there is no electric
field. We will count the rotation phase starting from
the particle crossing of the shock at $t=0$. Particle motion in $y$-
direction upstream relative to the shock is then given by $y=\rho\left[\sin\left(\phi+\omega_{c}t\right)-\sin\left(\phi\right)\right]-ut$,
where $\rho=v_{\perp}/\omega_{c}$ is the particle's gyroradius, $\omega_{c}$
- the relativistic gyrofrequency and the initial phase $-\pi/2\le\phi\le\pi/2$
is related to the upstream entry momentum as follows $\tan\phi=-\left(p_{x}/p_{y}\right)$
(see Fig.\ref{fig:Particles-bouncing-between}). It is convenient
to denote the rotation phase, at which the particle reenters the downstream
medium as $2\tau=\omega_{c}t_{\text{ret}}$. We write a transcendental
equation, $y\left(\tau\right)=0$, for $\tau$ in such a form as to
include only its root corresponding to the reverse crossing of the
shock, as the forward crossing occurs at $\tau=0$ automatically: 
\begin{equation}
\cos\left(\phi+\tau\right)\frac{\sin\tau}{\tau}=\frac{u\gamma}{p_{\perp}}=\frac{u}{v_{\perp}}\label{eq:MapTranscen}
\end{equation}
where $p_{\perp}=\sqrt{p_{x}^{2}+p_{y}^{2}}$. The momentum rotation
to the shock recrossing point $\left(p_{x}^{\prime},p_{y}^{\prime}\right)$
on the plane $\left(p_{x},p_{y}\right)$ upstream is given by
\begin{equation}
\begin{array}{cc}
p_{y}^{\prime}= & p_{\perp}\cos\left(\phi+2\tau\right)\\
p_{x}^{\prime}= & -p_{\perp}\sin\left(\phi+2\tau\right).
\end{array}\label{eq:Map2}
\end{equation}
The final step in the series of particle momentum transforms in eq.(\ref{eq:MapSeries})
is the same as in eq.(\ref{eq:Map1}) but with the replacement $u\to-u.$
To complete the acceleration cycle, we simply repeat the series of
transforms in eq.(\ref{eq:MapSeries}), applying them to the second
shock. Further aspects of the map we found particularly useful
for its generalizations are collected in Appendix \ref{sec:Appendix-B}. 

As stated earlier, we map an ingress angle, $0\le\hat{\alpha}\le\pi$,
that a particle makes with the shock surface on a plane perpendicular
to the magnetic field, Fig.\ref{fig:Particles-bouncing-between}.
Note that an egress angle, $\check{\alpha}$, is supplementary to
what is normally considered as a reflection angle, $\check{\alpha}=\pi-\alpha_{\text{refl}}$.
This notation is convenient since after turning to the second shock,
the ingress angle $\hat{\alpha}_{2}=\check{\alpha}_{1}$, provided
that the shocks are parallel to each other. More importantly, fixed
points (FPs) of the iterated map (periodic orbits) satisfy the condition,
$\check{\alpha}=\hat{\alpha}$, not the conventional, $\check{\alpha}=\hat{\alpha}_{\text{refl}}$,
as the particle-shock interaction is not elastic, except for the limit
$u\to0$. 

On applying the Lorentz transforms to a particle initial upstream
phase, $\tan\phi=-\left(p_{x}/p_{y}\right)$, and, similarly, to its
final phase, $\phi^{\prime}=\phi+2\tau$, we first link them with
the downstream angles $\hat{\alpha}$ and $\check{\alpha}$:
\begin{align}
\phi=-\tan^{-1}\left(\frac{\cos\hat{\alpha}\sqrt{1-u^{2}}}{\sin\hat{\alpha}+u/\hat{v}_{\perp}}\right),\label{eq:fiOfAlphaHat_}\\
\phi^{\prime}=-\tan^{-1}\left(\frac{\cos\check{\alpha}\sqrt{1-u^{2}}}{\sin\check{\alpha}-u/\check{v}_{\perp}}\right)\label{eq:fiOfAlphaHat}
\end{align}
These relations differ only by the sign in front of $u$, as they
should. 

The sequence of the momentum transforms shown in eq.(\ref{eq:MapSeries})
can be expressed in terms of the relevant angles, which simplifies
the iterated map. The physical basis of this simplification is that
after several iterations (especially in relativistic shocks, $u\approx1$),
and parallel fields upstream of both shocks, the particle energy is
mostly in perpendicular motion, $v_{\perp}\approx1$. Ideally, we
need to find a direct map $\hat{\alpha}\mapsto\check{\alpha}$ at
one shock, then repeat it at the second shock, which will give us
the full cycle of particle acceleration. The problem, however, is
that in obtaining the $\hat{\alpha}\mapsto\check{\alpha}$ map, we
first need to find $\phi$, from eq.(\ref{eq:fiOfAlphaHat_}),
solve the transcendental equation for $\tau$, eq.(\ref{eq:MapTranscen}),
and, finally, invert the relation in eq.(\ref{eq:fiOfAlphaHat}),
to map $\phi^{\prime}\mapsto\check{\alpha}$. The most difficult step
is the transcendental equation for the upstream rotation parameter,
$\tau$, that cannot be given in a closed form for an arbitrary $\tau$. 

To tackle the above problem, instead of solving the transcendental
equation (\ref{eq:MapTranscen}) for the unknown upstream rotation
phase $\tau$, we use it as a parameter in finding the map $\hat{\alpha}\mapsto\check{\alpha}$,
by invoking the following algorithm. First, we invert the relations
in eqs.(\ref{eq:fiOfAlphaHat_},\ref{eq:fiOfAlphaHat}) to find $\hat{\alpha}\left(\phi\right)$
and $\check{\alpha}\left(\phi+2\tau\right)$. Second, we substitute
the phase $\phi$ from 
eq.(\ref{eq:MapTranscen}) in these formulae,
thus obtaining two explicit functions of $\tau$: $\hat{\alpha}\left(\tau\right)$
and $\check{\alpha}\left(\tau\right)$. These functions represent
an explicit parametric map $\hat{\alpha}\mapsto\check{\alpha}$, that
uses $\tau$ as a transformation parameter.

To simplify algebra, we assume $v_{\perp}=1$ for the analysis and
will treat the general cases numerically. Indeed, if the magnetic
fields ahead of the two shocks are parallel to each other; only the
perpendicular component of momentum increases. Thus, even if the particle
reacceleration starts with $v_{\perp}\lesssim1$, it quickly (depending
on $u$) approaches $v_{\perp}=1$. In this limit, the map $\hat{\alpha}\to\check{\alpha}$
takes a much simpler form than in the case of arbitrary $v_{\perp}\lesssim1$.
By executing the above-described steps, we obtain the following parametric
map of the ingress to egress angle, $\hat{\alpha}\to\check{\alpha}$: 
\begin{align}
\hat{\alpha} & = \cos^{-1}\left\{ -\frac{\sin\left[\cos^{-1}\left(u\tau\csc\tau\right)-\tau\right]\sqrt{1-u^{2}}}{1-u\cos\left[\cos^{-1}\left(u\tau\csc\tau\right)-\tau\right]}\right\} & \nonumber \\
\check{\alpha} & = \cos^{-1}\left\{ \frac{\sin\left[\cos^{-1}\left(u\tau\csc\tau\right)+\tau\right]\sqrt{1-u^{2}}}{1-u\cos\left[\cos^{-1}\left(u\tau\csc\tau\right)+\tau\right]}\right\} & \label{eq:AngleMap}
\end{align}

The particle energy gain per upstream excursion can be obtained
from a Lorentz transformation in terms of the entry phase, $\phi$,
and rotation phase, $2\tau$. Similarly to the above transformation
for the angles $\alpha$, this result can be cast in terms of $\tau$: 
\begin{align}
\eta\equiv\frac{\check{\gamma}}{\hat{\gamma}} & =  \frac{1-uv_{\perp}\cos\left(\phi+2\tau\right)}{1-uv_{\perp}\cos\phi} & \nonumber \\
 & \approx\frac{1-u\cos\left[\cos^{-1}\left(u\tau\csc\tau\right)+\tau\right]}{1-u\cos\left[\cos^{-1}\left(u\tau\csc\tau\right)-\tau\right]} & \label{eq:EnGain}
\end{align}
\begin{figure*}
\includegraphics[viewport=60bp 0bp 355bp 181bp,scale=0.527]{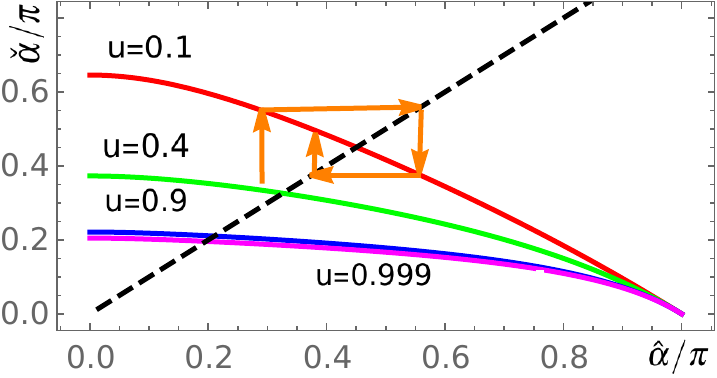}\includegraphics[viewport=-40bp 0bp 310bp 189bp,scale=0.365,viewport=0bp 0bp 411bp 217bp]{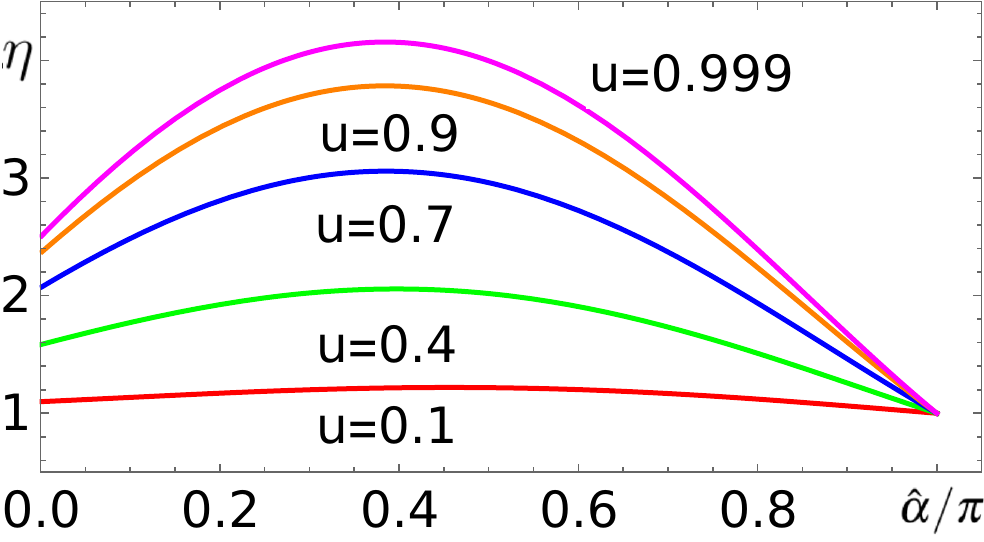}\includegraphics[viewport=-90bp -5bp 347bp 223bp,scale=0.31]{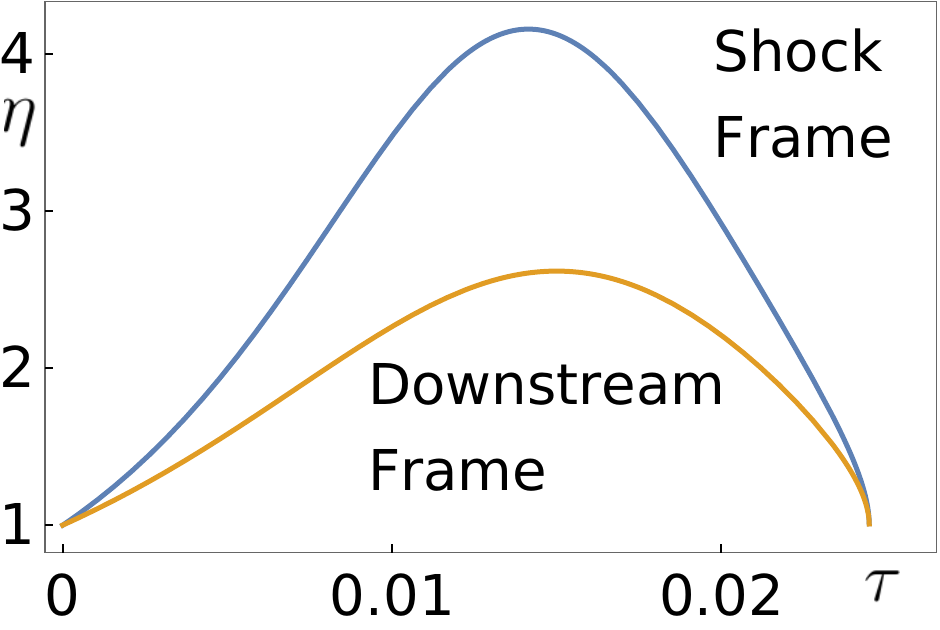}
\caption{Left Panel: Maps of ingress ($\hat{\alpha})$ to egress ($\check{\alpha}$)
angle after a particle excursion into the upstream medium, shown for
different shock speeds. The dashed line represents an identity map.
Map iterations (series of shock encounters) converging to a fixed
point (FP) are shown with the arrows (see text). Middle Panel: Energy
gain after one encounter with a shock as a function of the ingress
angle, shown in the shock frame for different shock speeds. Right
Panel: the same as in the middle panel for $u=0.9999$, but shown as a function of $\tau$
both in the shock frame and in a downstream frame that is moving at speed $u_{\text{d}}\approx u/3$ relative to the shock frame. \label{fig:AngleEnergyMaps}}
\end{figure*}
Again, the last expression is valid in the limit $v_{\perp}\to1$. 

Eqs.(\ref{eq:AngleMap}) and (\ref{eq:EnGain}) parametrically relate
the particle egress angle and energy gain after one shock encounter
to the particle ingress angle. Shown in Fig.\ref{fig:AngleEnergyMaps}
is the $\hat{\alpha}\mapsto\check{\alpha}$ map along with the dependence
of the energy gain $\eta\equiv\check{\gamma}/\hat{\gamma}$ upon $\hat{\alpha}$
for different shock velocities (left and middle panels). The arrows
on the left panel show the iterations of the map as they follow the
particle bouncing between the shocks. These iterations converge to
a single FP for which $\check{\alpha}=\hat{\alpha}$. The FP is stable,
since $K\equiv\left|d\check{\alpha}/d\hat{\alpha}\right|<1$ at $\check{\alpha}=\hat{\alpha}$.
Note that in the dynamical chaos theory, $K$ would be called a stochasticity
parameter. The meaning of this term is that if the FP becomes unstable,
$K>1$, iterations of $\alpha$ might become random, if $K$ exceeds
a stochasticity threshold, $K>K_{s}>1$. We will consider examples
of such dynamics in Sec.\ref{sec:Particle-Energy-Loss}, but
may conclude from Fig.\ref{fig:AngleEnergyMaps} that FPs associated
with relativistic shocks are more stable (smaller $K<1$) with a faster
convergence of iterations. Such shocks would need stronger perturbations
to the iterated map before it becomes unstable, i.e., to increase
$K>1$. This conclusion impacts the particle escape from the shock
system when caused by their stochastic scattering. This observation 
and the higher energy gain of relativistic shocks make them more 
promising in accelerating high-energy particles.

We further illustrate the dynamics of convergence to an FP  in Fig.\ref{fig:Convergence-of-iterations}.
Again, it is quantitatively different for nonrelativistic and
relativistic shocks. In the first case, the convergence is very slow,
taking several thousands of iterations to reach the FP when $u\lesssim10^{-3}$. Moreover, 
this number rapidly grows as $u$ decreases below this value. Conversely,
a pair of relativistic shocks bring a particle bouncing between the
shocks to a steady state after a few bounces. During early iterations, the interaction with nonrelativistic
shocks is similar to a conventional elastic reflection, whereby the
incidence angle equals to the reflection angle. 
 At this stage, particles hitting nonrelativistic shocks almost tangentially
may escape before reaching the FP. Asymptotically, however, both $\check{\alpha}$
and $\hat{\alpha}$ converge to a common angle $\beta=$$\hat{\alpha}=\check{\alpha}<\pi/2$,
with $\beta\to\pi/2$, for $u\to0$. The dependence of this angle
upon $u$ is shown in Fig.\ref{fig:EnGainAndAnglesVs-u}. As $u$
grows, the angle $\beta$ decreases, and the deviation from a specular
reflection becomes more pronounced. After reaching the steady state,
the particle's return path downstream aligns with its forward
path. In this state, the reflection angle becomes supplementary rather
than equal to the incidence angle if one uses the terminology of
ray reflection from a surface. 

The energy gain per cycle at the FP is very high but somewhat lower
than its absolute maximum value that exceeds four, Fig.\ref{fig:AngleEnergyMaps}.
Note that this maximum value ($\eta_{\text{max}}\approx4.15$) is
higher than that found for a similar scatter-free particle
interaction with ultrarelativistic shocks by \cite{GallantAcht99,Achterberg2001},
($\eta_{\text{max}}\approx2.62$). The reason for the difference is
that these authors considered a single shock acceleration and the
result was given in the downstream frame that moves at $u_{\text{d}}\approx u/3$
relative to the shock frame, in which we present all our pertinent results
to the double-shock acceleration. To transform the energy gained from
the shock to the moving downstream frame, it is sufficient to replace $u$
in front of $\cos\left(\dots\right)$ in eq.(\ref{eq:EnGain}) by
a relative velocity between the upstream and downstream frames $u_{\text{rel}}=\left(u-u_{\text{d}}\right)/\left(1-u_{\text{d}}u\right)\approx2u/\left(3-u^{2}\right)$.
The difference in the energy gain between the two reference frames
is shown in the right panel of Fig.\ref{fig:AngleEnergyMaps}. 
\section{Properties of the Iterated Map}\label{sec:Properties-of-Map}

\begin{figure*}
\includegraphics[scale=0.71]{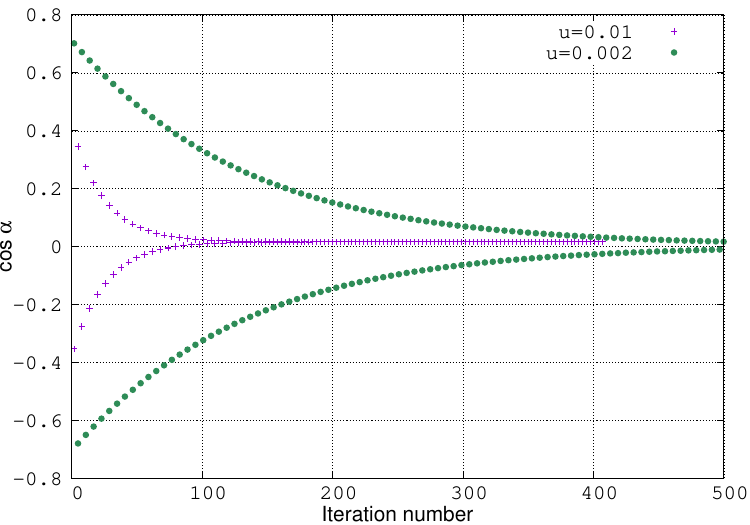}\includegraphics[scale=0.71]{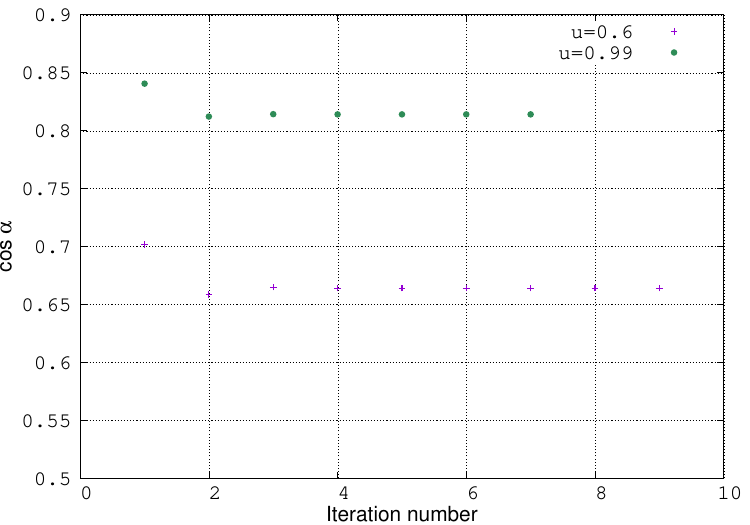}
\caption{Particle bouncing between nonrelativistic (left panel) and relativistic
(right panel) shocks. Particles keep flipping the cosine of ingress
angle for much longer time in nonrelativistic than relativistic
shocks before they attain a constant angle regime corresponding to
an FP. \label{fig:Convergence-of-iterations}}
\end{figure*}
 Fixed points (FPs) are invaluable in studying iterated maps. However,
the FP described in the previous section merits clarification. First,
lossless particles gain momentum (Fig.\ref{fig:AngleEnergyMaps})
each time they visit upstream media because of the motional electric
field. Hence, the particle trajectory never returns to the same point
in the phase space, meaning there are no FPs in the usual
sense. Nevertheless, the map of angular variables might reach an FP
when $p_{\parallel}/p\to0$. If the iterations start from an arbitrary
momentum, this limit is reached when $v_{\perp}\to1$. We keep $v_{\perp}$ arbitrary to handle the limit
$u\to1$ properly. According to eq.(\ref{eq:MapTranscen}), for example, $\cos\phi\to1$
in this case, since $\tau\to0$ is required in the limit $u/v_{\perp}\to1$,
meaning such particles merely touch the shock surface before
returning downstream.

Second, the orbit may drift along the shock surface, which occurs,
e.g., when the shock velocities are not equal. In this case, the FP
will be related to the ingress and egress \emph{angles}, while the
drift will proceed with the same step in each cycle. It can still
be considered as a periodic orbit, modulo the step. In this consideration,
however, the increase of the particle gyroradius during its motion upstream
is neglected compared to the orbit displacement occurring between the shocks.

As a side note, our system is similar to billiards, e.g., \cite{Lichtenberg1983,Loskutov2007,ZaslavHamChaos07},
with the exception that reflections off the upstream medium are inelastic.
Drifts also occur in infinite billiards. They do not matter in either
of these systems as long as the coordinate in the drift direction
is ignorable and the drift can be considered modulo the step, as we
noted. For finite shocks, it determines the particle escape time. 

\begin{figure}%
\includegraphics[scale=0.69]{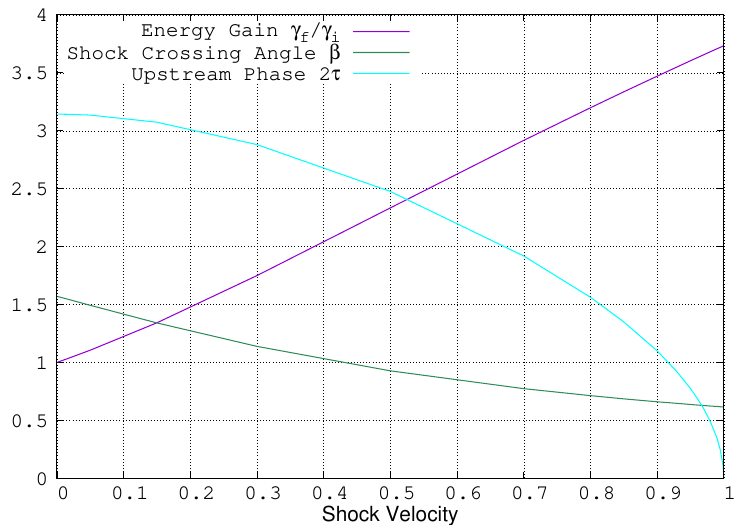}
\caption{Particle kinematic parameters as functions of the shock speed, $u$.
Ingress angle, $\beta$, upstream rotation phase, $2\tau$, and the
energy gained after each upstream excursion are shown for a steady
state (fixed point of the iterated map).\label{fig:EnGainAndAnglesVs-u}}
\end{figure}%

A full-cycle FP (after a particle has visited both shocks) can be
defined, e.g., as $\hat{\alpha}_{1}=\check{\alpha}_{2}$ (the dashed
line orbit in Fig.\ref{fig:Particles-bouncing-between}, while the
indices 1 and 2 here refer to the shock numbers). However, considering
a special case of parallel shocks of the same speed $u$, and taking
into account that the map $\hat{\alpha}_{1,2}\mapsto\check{\alpha}_{1,2}$
are injective (one-to-one), the above FP condition requires $\hat{\alpha}_{1,2}=\check{\alpha}_{1,2}$,
because $\hat{\alpha}_{2}=$$\check{\alpha}_{1}$ and $\hat{\alpha}_{1}=$$\check{\alpha}_{2}$.
Therefore, it is sufficient to obtain only a single-shock FP, defined
as $\check{\alpha}=\hat{\alpha}$. To find this FP, one may use the
parametric relations in eqs.(\ref{eq:AngleMap}) for these angles.
However, as this is the simplest FP, we may directly use the condition
that all paths of a particle between the shocks are parallel to each
other. Using the notations introduced in Sec.\ref{sec:IteratedMap},
we write the following identity for the particle velocity
\[
\frac{\hat{v}_{y}}{\hat{v}_{x}}=\frac{\check{v}_{y}}{\check{v}_{x}}.
\]
Now we can express the 'hat' (in) and 'check' (out) variables, given
here in the downstream frame (which, in fact, is the shock rest frame),
through their upstream values at the lower and upper shock, respectively.
Using the Lorentz transformations and notations used for the map in
eq.(\ref{eq:MapSeries}), the last relation rewrites
\[
\left(v_{y}-u\right)v_{x}^{\prime}=\left(v_{y}^{\prime}-u\right)v_{x}.
\]
Note, that the common factors $1-uv_{y}$ and $\sqrt{1-u^{2}}$ in
velocity transformations have canceled out. To reach the FP, the rotation phases
$2\tau$ in the transform given by eq.(\ref{eq:Map2}) must be
the same for both shocks. Hence, the last relation rewrites
\[
\sin\left(\phi+2\tau\right)\left[v_{\perp}\cos\phi-u\right]=\sin\phi\left[v_{\perp}\cos\left(\phi+2\tau\right)-u\right]
\]
and can be transformed to the following simple form 
\[
\cos\tau=\frac{u}{v_{\perp}}\cos\left(\phi+\tau\right)
\]
The last relation can be used to express $\phi$ through 
$\tau$ explicitly. To find the fixed point, we add to this equation the general
relation between $\phi$ and $\tau$ given by eq.(\ref{eq:MapTranscen}).
We thus obtain the following system of two equations for $\phi$ and
$\tau$:
\begin{equation}
\cos\tau=\frac{u}{v_{\perp}}\cos\left(\phi+\tau\right)=\left(\frac{u}{v_{\perp}}\right)^{2}\frac{\tau}{\sin\tau}\label{eq:FPandTranscen}
\end{equation}

The transcendental equation in eq.(\ref{eq:MapTranscen}) (the second
relation above) for the rotation phase $2\tau$ is now supplemented
by the first relation, from which we can find both the particle entry
phase $\phi$ and the rotation phase $2\tau=\omega_{c}t$, where $t$
is the time a particle spends upstream of a shock. Note, that $t$
is not the same in the energy lossless case; only $\tau$ is. These
expressions allow us to represent the particle energy gain per one
upstream excursion given in eq.(\ref{eq:EnGain}) in terms of its
upstream starting phase, $\phi$, or the shock incidence angle at
the FP $\beta=\hat{\alpha}=\check{\alpha}<\pi/2$. For the first task,
we rewrite the general relation for the energy gain given by eq.(\ref{eq:EnGain})
in a more symmetric form
\[
\eta\equiv\frac{\check{\gamma}}{\hat{\gamma}}=\frac{1-uv_{\perp}\cos\left(\psi+\tau\right)}{1-uv_{\perp}\cos\left(\psi-\tau\right)}
\]
where $\psi\equiv\phi+\tau$. By expanding $\cos\left(\psi\pm\tau\right)$
and applying the left relation in eq.(\ref{eq:FPandTranscen}), again,
in the limit $v_{\perp}\to1$ the last result rewrites as follows
\[
\eta=\frac{\sin\tau+u\sin\psi}{\sin\tau-u\sin\psi}
\]
We may multiply the numerator and denominator by denominator and use
the left relation in eq.(\ref{eq:FPandTranscen}) again. This yields
a relation for $\eta$ at the FP that depends only on the upstream
entrance phase $\phi$:
\begin{equation}
\frac{\check{\gamma}}{\hat{\gamma}}=\frac{1-u^{2}}{1-2u\cos\phi+u^{2}}\label{eq:EnGainFP}
\end{equation}
To obtain the energy gain as a function of the FP ingress angle $\beta=\hat{\alpha}=\check{\alpha}$,
it is sufficient to express $\cos\phi$ from the general relation
between $\phi$ and $\hat{\alpha}$, given in eq.(\ref{eq:fiOfAlphaHat_}).
It can be written as follows: $\cos\phi=\left(u+\sin\beta\right)/\left(1+u\sin\beta\right)$.
After simple algebra, we finally obtain
\[
\frac{\check{\gamma}}{\hat{\gamma}}=\frac{1+u\sin\beta}{1-u\sin\beta}.
\]

The angles $\phi$ and $\beta$ come from the solution of the transcendental
equation for the upstream rotation parameter $\tau$ in eq.(\ref{eq:FPandTranscen}).
In general, it needs to be solved numerically, but simple analytic
solutions can be obtained for $u\ll1$ and $u\approx1$ (we set $v_{\perp}=1$
here, as remarked earlier). For a nonrelativistic shock, $u\ll1$,
we find $\tau\approx\pi\left(1-u^{2}\right)/2\approx\pi/2$ and $\phi\approx\pi u/2$,
so that $\eta\equiv\check{\gamma}/\hat{\gamma}\approx1+2u$, which
is the energy, gained after an elastic head-on collision with a heavy
nonrelativistic particle (magnetic field frozen in the upstream plasma).
In the case of $u\approx1$, we obtain $\tau\approx\sqrt{3\left(1-u\right)}$
and $\phi\approx-\left(\sqrt{3}-1\right)\sqrt{1-u}$. The energy gain
at $u\to1$ is:
\begin{equation}
\frac{\check{\gamma}}{\hat{\gamma}}=\frac{1-u^{2}}{\left(1-u\right)^{2}+\phi^{2}}=\frac{2}{\left(\sqrt{3}-1\right)^{2}}\approx3.73\label{eq:EnGainFP-1}
\end{equation}
From eq.(\ref{eq:EnGainFP}) we can also extract the FP angle in this
limit, $\beta\approx\sin^{-1}\left(1/\sqrt{3}\right)\approx35.3^{\circ}$.

The FP ingress angle, $\beta$, upstream rotation phase, $2\tau$,
and the energy gain after each upstream excursion are shown in Fig.\ref{fig:EnGainAndAnglesVs-u}.
We have already discussed these quantities at the end of Sec.\ref{sec:IteratedMap}.
We merely add here that in a steady state, there is no drift along
the shock face, provided that the shock speeds are equal. However,
particles drift along the shock during the initial phase of shock-particle interaction, especially in the nonrelativistic case. Besides a
stronger FP stability, as mentioned earlier, the relativistic shocks
are more efficient in particle trapping. Hence, a particle's chance
for an early escape because of the pre-FP drift is diminished considerably
in relativistic shocks (cf. Fig.\ref{fig:Convergence-of-iterations}).

\section{Fate of Accelerated Particles}\label{sec:Particle-Energy-Loss}

In the previous section, we considered the acceleration of particles
trapped between two identical shocks with their fronts and magnetic
fields ahead of them being strictly parallel. We have
found that the particle orbit asymptotically enters and exits each
shock at the same angle. Apart from the growing energy, and consequently
increasing length of the upstream part of the orbit, we can still
call it a periodic orbit, or an FP of the iterated map for a single-shock
(half-cycle), rather than a two-shock (full-cycle) particle acceleration.
In the latter case, the periodicity is more general in double-shock
systems. Nevertheless, the usefulness of the half-cycle periodicity
will be obvious when we consider a period-doubling bifurcation. We
also note that if the ingress and egress angles were not equal
but the same at each shock, the solution could still be periodic in
terms of a fixed pair of angles. However, particles would be drifting
along the shock front, as shown in Fig.\ref{fig:Particles-bouncing-between}
by the dashed lines. We also know from the previous section that an
FP with unequal ingress-egress angles (and thus the particle drift)
is not possible if the shock velocities are the same. It is natural
to expect such drift when the shock speeds are different, which we
quantify in the next subsection. The significance of the drift is
that particles eventually escape because of it and the shock size determines their maximum energy.

The FP conditions considered in the previous section may not be realistic
enough to account for particle escape properly. However, the energy
gain after each cycle is high enough for the acceleration mechanism
to be very efficient even if particles escape early. In highly relativistic
shocks, in particular, the energy gain per cycle  ({\it i.e.}, two-shock interaction) reaches a factor of
fourteen near an FP. So, even a few cycles may increase the particle
energy to the point when the question of how it stops growing becomes
relevant regardless of the escape. For the leptons, the synchrotron
radiation or inverse Compton scattering, especially in a high photon
flux environment in binary stars, e.g.~\cite{2006MNRAS.372..801P}, limit
the maximum energy. However, the acceleration limit for the hadrons
is likely set by their escape, which deserves a closer look. 

Apart from escaping due to the unequal shock speeds, particles might
also escape if one or both shocks are rippled. An FP may still exist
but lose stability (if parameter $K>1$, introduced in Sec.\ref{sec:IteratedMap}).
Particles will drift or diffuse along the shock front and eventually
escape. We will also consider this scenario, but include particle
energy losses as a free parameter. In this case, the FP becomes absolute
because the particle energy does not change between cycles, as the losses absorb the
energy gain. Finally, if the magnetic field
directions upstream of the shocks are not the same, particles might
drift along the shock surfaces. We address the above three escape
mechanisms below.

\subsection{Deterministic Escape Scenario\label{subsec:Deterministic-Escape-Scenario}}

Let us now apply the iterated map formulas for ingress and egress
angles, derived for a single shock in eq.(\ref{eq:AngleMap}), to
a pair of shocks of different speeds, i.e., $u_{1}\neq u_{2}.$ A
simple FP with a particle that bounces between two fixed points on
the opposite shock surfaces, considered in the previous section, is
impossible in this case. The question is how fast the particle
will drift and escape, depending on $u_{1}$ and $u_{2}$.

The FP condition per one cycle (two particle-shock collisions in a
row) can be written as follows: $\check{\alpha}_{1}=\hat{\alpha}_{2}$
and $\check{\alpha}_{2}=\hat{\alpha}_{1}$ (shock -1 egress angle
equals shock-2 ingress angle and vice versa). To shorten the notation,
we will use the cosines of the angles and the upstream phases $\phi_{1,2}$
used earlier in Sec.\ref{sec:IteratedMap}. We will also use the angular
velocity representations $\beta_{1,2}=\cos^{-1}u_{1,2}$, for convenience; note therefore the difference with respect to $\beta=\hat{\alpha}=\check{\alpha}$ introduced in the previous Section.
The above FP conditions for the angles read:
\begin{align}\cos\check{\alpha}_{1}=\frac{\sin\left(\beta_{1}\right)\sin\left(2\tau_{1}+\phi_{1}\right)}{1-\cos\left(\beta_{1}\right)\cos\left(2\tau_{1}+\phi_{1}\right)}=\nonumber\\
\cos\hat{\alpha}_{2}=\frac{\sin\left(\beta_{2}\right)\sin\left(\phi_{2}\right)}{\cos\left(\beta_{2}\right)\cos\left(\phi_{2}\right)-1}\label{eq:FP1}\\
\cos\check{\alpha}_{2}=\frac{\sin\left(\beta_{2}\right)\sin\left(2\tau_{2}+\phi_{2}\right)}{1-\cos\left(\beta_{2}\right)\cos\left(2\tau_{2}+\phi_{2}\right)}=\nonumber\\
\cos\hat{\alpha}_{1}=\frac{\sin\left(\beta_{1}\right)\sin\left(\phi_{1}\right)}{\cos\left(\beta_{1}\right)\cos\left(\phi_{1}\right)-1}\label{eq:FP2}
\end{align}

As in Sec.\ref{sec:IteratedMap}, $2\tau_{1,2}$ denote the upstream
rotation phases at each shock. By adding two equations for the rotation
phases, given by eq.{[}\ref{eq:MapTranscen}{]}, 
\begin{equation}
\cos\left(\phi_{1,2}+\tau_{1,2}\right)=\cos\beta_{1,2}\tau_{1,2}/\sin\tau_{1,2},\label{eq:FP34}
\end{equation}
we now have a system of four equations for $\phi_{1,2}$ and $\tau_{1,2}$.

In the case of arbitrary $u_{1}$ and $u_{2}$, the above equations
should be solved numerically. If $u_{1}-u_{2}\sim1$, particles will
drift along the shocks at a significant fraction of the speed of light.
They may still gain much energy if at least one of the shocks is relativistic.
If both shocks are relativistic so that $\left|u_{1}-u_{2}\right|\ll1$,
we can use this quantity as a small parameter to solve the above problem
analytically. The advantage here is that the analytic solution explicitly
shows the escape efficiency as a function of shock velocities. Besides,
this case is the most favorable for acceleration, as it gives the
highest ratio of the energy gain to the escape rate. 

\begin{figure}%
\includegraphics[viewport=30bp 0bp 640bp 480bp,scale=0.45]{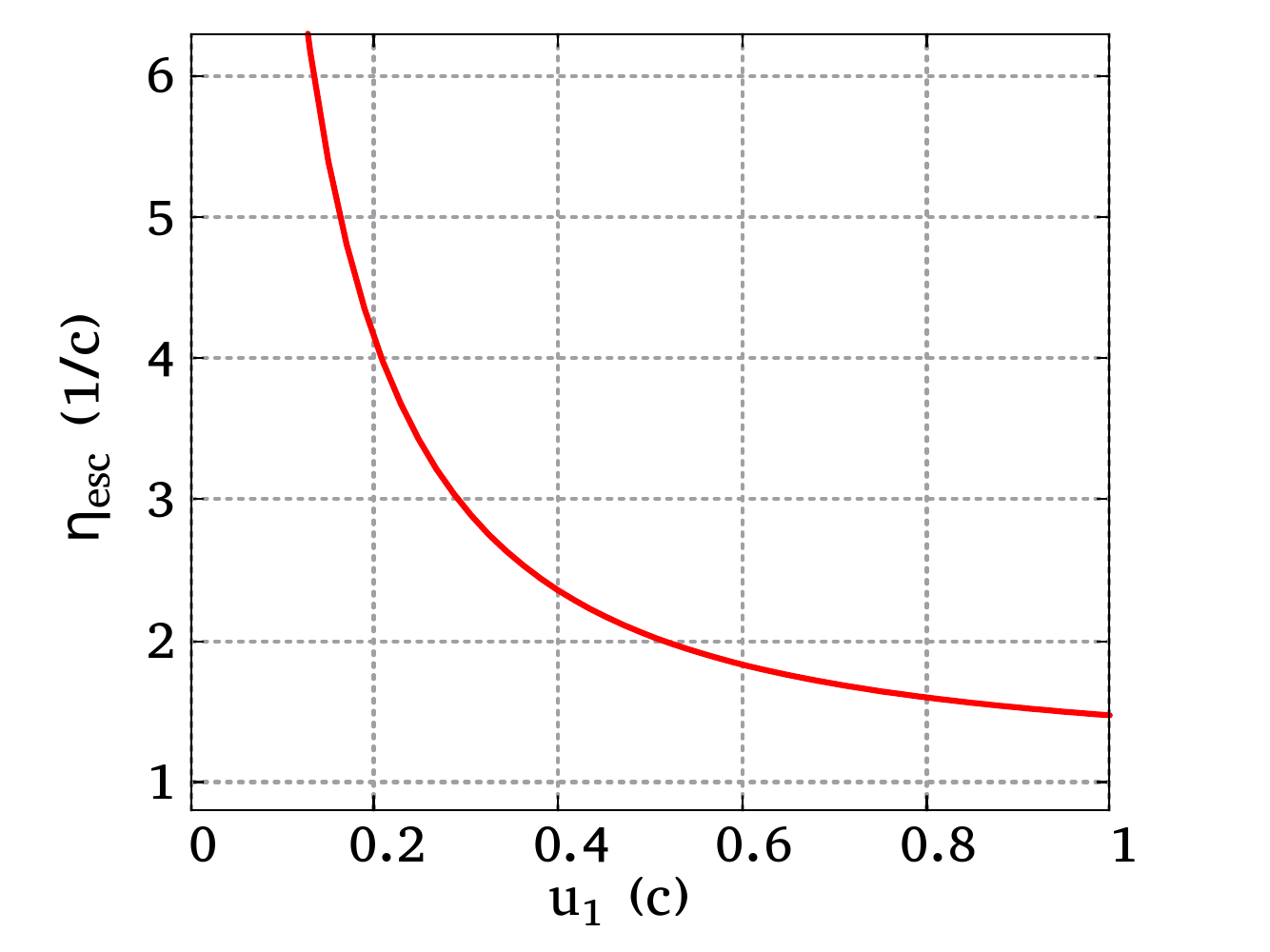}
\caption{Normalized particle escape rate vs the speed of one
of two shocks, $u_{1}$, provided that $\left|\Delta u\right|\equiv\left|u_{1}-u_{2}\right|\ll1$. The units of measurement are shown in parentheses.
\label{fig:Normalized-particle-escape}}
\end{figure}%

In the case $\beta_{2}=\beta_{1}$, that is $u_{1}=u_{2}$, the solution
of the above system coincides with that we have found in the previous
section. It is given by eqs.(\ref{eq:FPandTranscen}). Expanding eqs.(\ref{eq:FP1}-\ref{eq:FP34})
in small $\left|\beta_{2}-\beta_{1}\right|\ll1$, after some algebra
we find
\begin{align}
\Delta\cos\alpha\equiv\cos\check{\alpha}_{1}-\cos\hat{\alpha}_{1}=\nonumber\\
\frac{\left(\beta_{2}-\beta_{1}\right)\sec\beta_{1}\cot\tau\csc\tau\left[3\cos\left(2\beta_{1}\right)-2\cos\left(2\tau\right)-1\right]}{2\left[\cos\left(2\tau\right)-\cos\left(2\beta_{1}\right)\right]\left(1-\sec^{2}\beta_{1}\cos^{2}\tau\right)^{-1/2}}\label{eq:DelCos}
\end{align}
where $\tau=\tau\left(\beta_{1}\right)$ is the zeroth order solution
for the FP (already found in the previous section), obtained from the following
equation: $\sin\left(2\tau\right)=$$2\tau\cos^{2}$$\beta_{1}$.
Using the standard notation for the function $\ensuremath{\sin\left(x\right)/x}\equiv\text{sinc}\left(x\right)$,
we can substitute $\tau=\text{\ensuremath{\frac{1}{2}\text{sinc}^{-1}\left(\cos^{2}\beta_{1}\right)}}$
in eq.(\ref{eq:DelCos}). Note that we have not used the second of
the two conditions, $\beta_{2}\approx\beta_{1}\approx1$, as yet,
so our treatment is also valid for nonrelativistic shocks as long
as $\beta_{2}\approx\beta_{1}$. 

The particle escape rate can be characterized by a normalized displacement
per cycle (two consecutive particle-shock collisions)
\[
\Delta x/d\approx\Delta\cot\alpha\equiv\cot\check{\alpha}_{1}-\cot\hat{\alpha}_{1},
\]
where $d$ is the gap between the shocks. We infer from eq.(\ref{eq:DelCos})
that this quantity is proportional to a (small) difference between
the shock speeds, $\Delta u=u_{1}-u_{2}\ll1$. Therefore, for a better
sense of particle escape rate, depending on the shock speed, we divide
the above quantity by $\Delta u$ and introduce the following normalized escape rate:
$$\eta_{\rm esc}=\frac{\Delta x}{d\Delta u}.$$
 Given $\tau=\tau\left(\beta_{1}\right)$,
defined earlier, our normalized escape rate will depend only on $u_{1}=\cos\beta_{1}$:
\begin{equation}
\eta_{\rm esc}=\frac{\sin(2\tau)\left[3\cos\left(2\beta_{1}\right)-2\cos\left(2\tau\right)-1\right]}{2\sin\left(2\beta_{1}\right)\left[\cos\left(2\beta_{1}\right)-\cos\left(2\tau\right)\right]\left[1-\sec^{2}\beta_{1}\cos^{2}\tau\right]}\label{eq:EscRate}
\end{equation}
This escape rate is shown in Fig.\ref{fig:Normalized-particle-escape}.
It is seen that in the case of relativistic shocks, particle escape
is indeed significantly suppressed compared with nonrelativistic shocks. Even for
a wide intershock space, $d\sim L$, where $L$ is the shock size,
it would take $\sim\Delta u^{-1}\gg1$ acceleration cycles before
a particle escapes the double-shock system. Given the high energy
gain at each cycle, the total energy gain, $\gamma_{\text{f}}/\gamma_{\text{i}}$
is very high indeed, $\ln\left(\gamma_{\text{f}}/\gamma_{\text{i}}\right)\sim1/\Delta u\gg1$.

\subsection{Route to Chaos and Escape}\label{sec:Route-to-Chaos}

Up to now, we have considered an ordinary particle acceleration with a
continuously growing energy. If energy losses are included, and a particle
reflection angle is perturbed, the periodic particle orbit might become
chaotic. As we remarked earlier, the perturbation does not have to
be random. A sinusoidal shock corrugation may trigger chaos. 

Due to the combination of the above two factors, an initial phase space
volume expands in one direction and contracts in the other. Let us select a 
plane region on a projection of the phase space. Since the total number
of particles is conserved, the area of this region will remain the
same under the mapping (area-preserving map). At the same time, its form becomes increasingly
complicated because it is stretched, contracted, and folded by the map. The dynamical chaos often commences (including the case
considered below) as a series of period-doubling bifurcations of a
particle orbit \cite{Feigenbaum1979,Feigenbaum1980} under a gradually
changing control parameter. The first such bifurcation is formally
similar to a sequence of alternating ingress-egress angles considered
in the preceding subsection. This time, the reason is the
shock corrugation rather than different shock velocities. 

\begin{figure*}
\includegraphics[viewport=0bp 6.54804bp 503bp 368bp,scale=0.328]{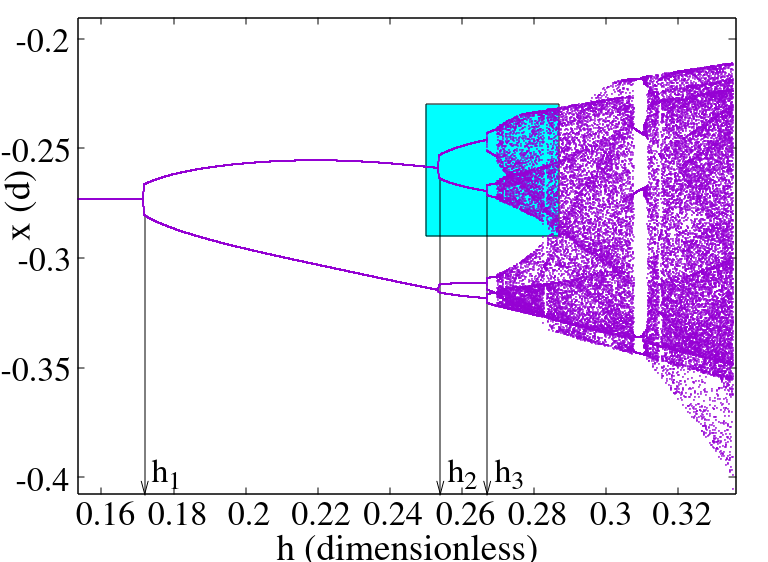}\includegraphics[viewport=13.08271bp 0bp 696bp 503bp,scale=0.234]{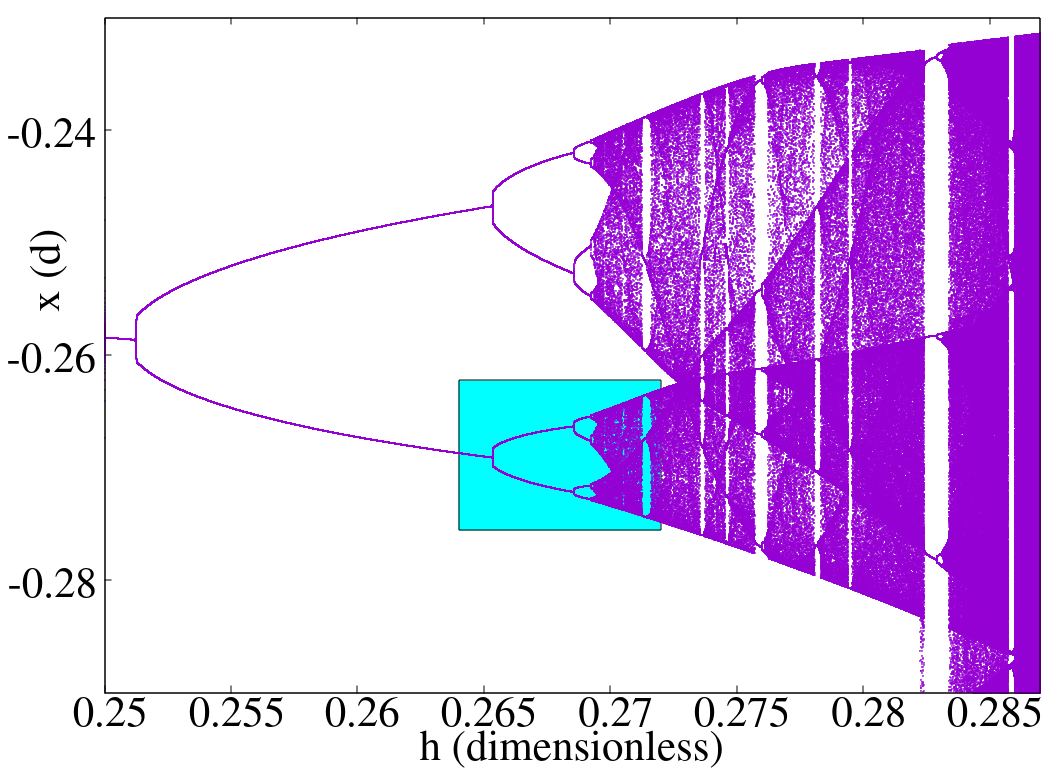}\includegraphics[viewport=0bp 3.27014bp 645bp 414bp,scale=0.287]{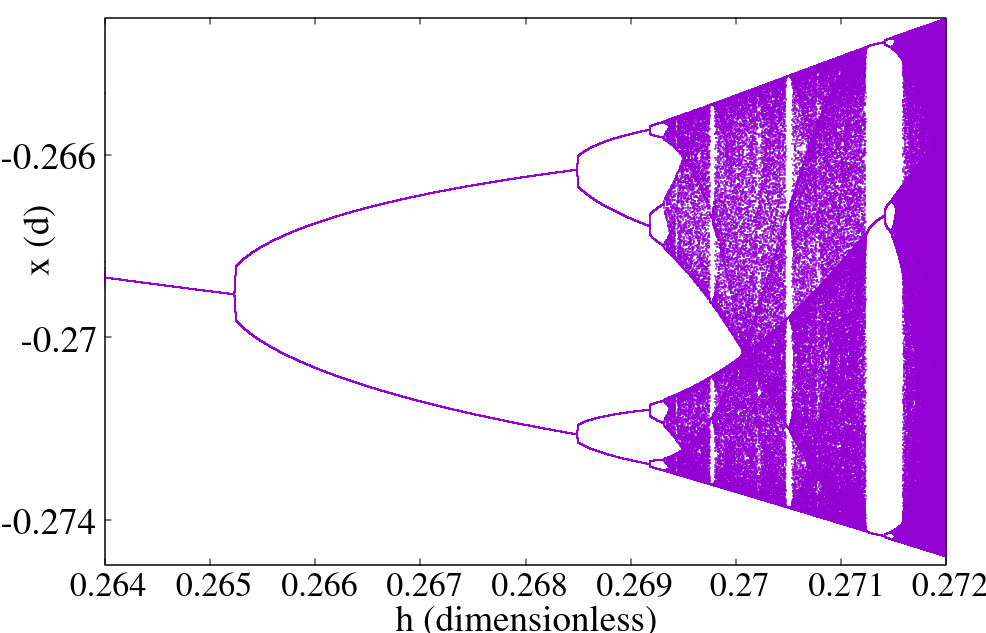}
\caption{Left: Iterated map for particle intersections at $x$ (vertical axis)
in eq.(\ref{eq:Corrug-d}) with the unperturbed shock. The second
shock is perturbed according to eq.(\ref{eq:Corrug-d}) with  the amplitude $h$
shown on the horizontal axis. Period doubling bifurcations are traced
by scanning the $x$ values of the map when $h$ changes within
its range. Middle: zoom into the rectangle in the left panel. Right:
zoom into the rectangle in the middle panel. \label{fig:PerDoublBif} }
\end{figure*}

To illustrate the onset of stochasticity in particle dynamics, let
us impose a sinusoidal perturbation on the surface of one of the shocks.
In effect, the gap $d$ between the shocks now varies with $x$ (coordinate
along the shock front but perpendicular to the magnetic field upstream
of each shock). The magnetic field is in the same direction upstream
of both shocks. Our iterated map operates on the ingress-egress angles. 
At the same time, the intershock parts of trajectories are rectilinear, and the
upstream parts of the particle displacement along the shock front
are neglected. We can normalize the average distance between the shocks
to unity, $\bar{d}=1$. The particle physical displacement along the
shock will be measured in these units. We define the gap $d$ as
follows:
\begin{equation}
d\left(x\right)=1+\frac{h}{k}\sin\left(kx\right),\label{eq:Corrug-d}
\end{equation}
where $h$ is an amplitude- and $k$ - a wavenumber of corrugation.
If we assume that the lower shock coincides with the plane $y=0$,
then $d\left(x\right)$ is the $y$- coordinate of the local position
of the upper shock. Assuming also that $h/k\ll1$ we can simplify
the map by neglecting the perturbation of the intershock distance compared with 
the perturbations of particle ingress-egress
angles. 

While crossing the intershock space, particles incur energy losses
that allow for a genuine FP of the iterated map. This FP includes not only
the angles of a particle trajectory but also the particle energy.
We introduce the losses in a simple form, assuming that they
are quadratic in energy (as e.g., synchrotron losses). We then include the
following transformation of the particle $\gamma$- factor after it
crosses the gap: $1/\gamma\mapsto1/\gamma+1/\gamma_{c}$, where $\gamma_{c}$
is an energy cut-off associated with the losses. This rule follows
from our assumption that the losses are quadratic: $d\gamma/dt\propto-\gamma^{2}$, 
$d\gamma^{-1}/dt\propto const$.
Strictly speaking, the loss-parameter $\gamma_{c}$ is proportional
to the length of the intershock trajectory. Nevertheless, focusing on the dynamics
near a fixed point, we can keep $\gamma_{c}$ constant.

A convenient Poincar\'e section variable to describe the map dynamics
is the cotangent of the ingress angle at the unperturbed shock. For $h/k\ll1$,
it approximately corresponds to a shift in the particle coordinate
$x$ as it moves from the perturbed to unperturbed shock. The next
value of $x$ to use in eq.(\ref{eq:Corrug-d}) is then calculated
by adding to the unperturbed shock value of $x$ a cotangent of the
egress angle. We will map only one of these $x$-values, namely the
one at the unperturbed shock, as a marker of a full acceleration cycle. 

The particle interaction with the lower shock remains the same as
described in Sec.\ref{sec:IteratedMap}. Before applying the angle
transformation formulae (\ref{eq:AngleMap}) to the upper shock we rotate
the $xy$ plane by the shock angle $\sigma=\tan^{-1}\left[h\cos\left(kx\right)\right]$
and then rotate it back after the particle left the perturbed shock.
Equivalently, one can subtract (add) this angle to a respective ingress
angle at the upper (lower) shocks. This equivalence reflects an additive
group property of the map elements. A general group representation
of the map that allows the inclusion of additional transformations in the form of respective matrices as elements of additive groups is given in
Appendix \ref{sec:Appendix-B}.

Shown in Fig.\ref{fig:PerDoublBif} is a scan in the modulation amplitude,
$h$, of $x$- intercepts (vertical axis) of a particle orbit with
a flat shock while the other shock is modulated according to eq.(\ref{eq:Corrug-d}).
The scan is generated as follows. Each time the modulation amplitude,
$h$, is increased by a small amount, the map is iterated until it
converges to a finite sequence of $x$- points that repeat periodically.
The scan starts from a subcritical value of $h$ when the iterated
map converges to a single point, which is an FP of the map, as shown
in the left panel. This is regarded as a period-one orbit. When $h$
reaches the first critical value, $h_{1}\approx0.1711$, the map abruptly
splits into two branches, constituting a period-two orbit of the map.  The
further increase of $h$ leads, at first, to a gradual divergence
of these branches until each of them splits into two points again at
$h=h_{2}\approx0.2532$. The period-doubling bifurcations continue
\emph{ad infinitum} with a rapidly decreasing separation between the
subsequent branching points $\left|h_{n+1}-h_{n}\right|\to0$ as $n\to\infty$.
The sequence $\left\{ h_{n}\right\} $ thus converges to a finite
$h=h_{\infty}$. At this point, the period is infinite, and the sequence
$\left\{ x_{k}\right\} _{k=1}^{\infty}$ becomes chaotic densely covering
a finite interval $\Delta x$. Recall that the chaos is caused by
a perfectly deterministic (single-mode) perturbation of the surface
of one of the two shocks. This phenomenon was studied in the seminal
works of Feigenbaum \cite{Feigenbaum1979,Feigenbaum1980}.

One might infer from the left panel of Fig.\ref{fig:PerDoublBif}
that the map becomes chaotic already after four-five bifurcations.
The iterations appear to cover subintervals of proliferated
overlapped branches densely. This would be a deceptive inference caused by
the limited resolution of the plot shown in the left panel. Indeed, by
zooming into the ``chaotic'' area (see middle and right panels)
we observe that the period-doubling bifurcations continue in a morphologically
self-similar fashion. The number of branches tends to infinity
when $h$ tends to some finite value, $h\to h_{\infty}-0$, where
chaos sets in. Nonetheless, a backward bifurcation occurs at
$h=h_{\infty}\approx0.2694$, which is best seen in the right panel
of Fig.\ref{fig:PerDoublBif} as the left-most narrow gaps in the
chaotic ``sea'' of the map. The number of branches to which each
chaotic area collapses is three. They persist until the next period-doubling 
occurs on each of these regular branches according to the
above-described period-doubling scenario. 

On a practical note, the narrowness (measure zero sets) of chaotic
regions in the control parameter, $h$, is also deceptive. If we add
a weak random noise to the map at each step, the intervals of $h$
with chaotic dynamics will broaden, e.g., \cite{schuster2006deterministic}.
A crucial conclusion from this result is that the main
parameters of chaotic dynamics, such as the particle diffusivity along
the shock face, are set by deterministic chaos.

\begin{figure*}
\includegraphics[viewport=0bp 6bp 951bp 593bp,scale=0.275]{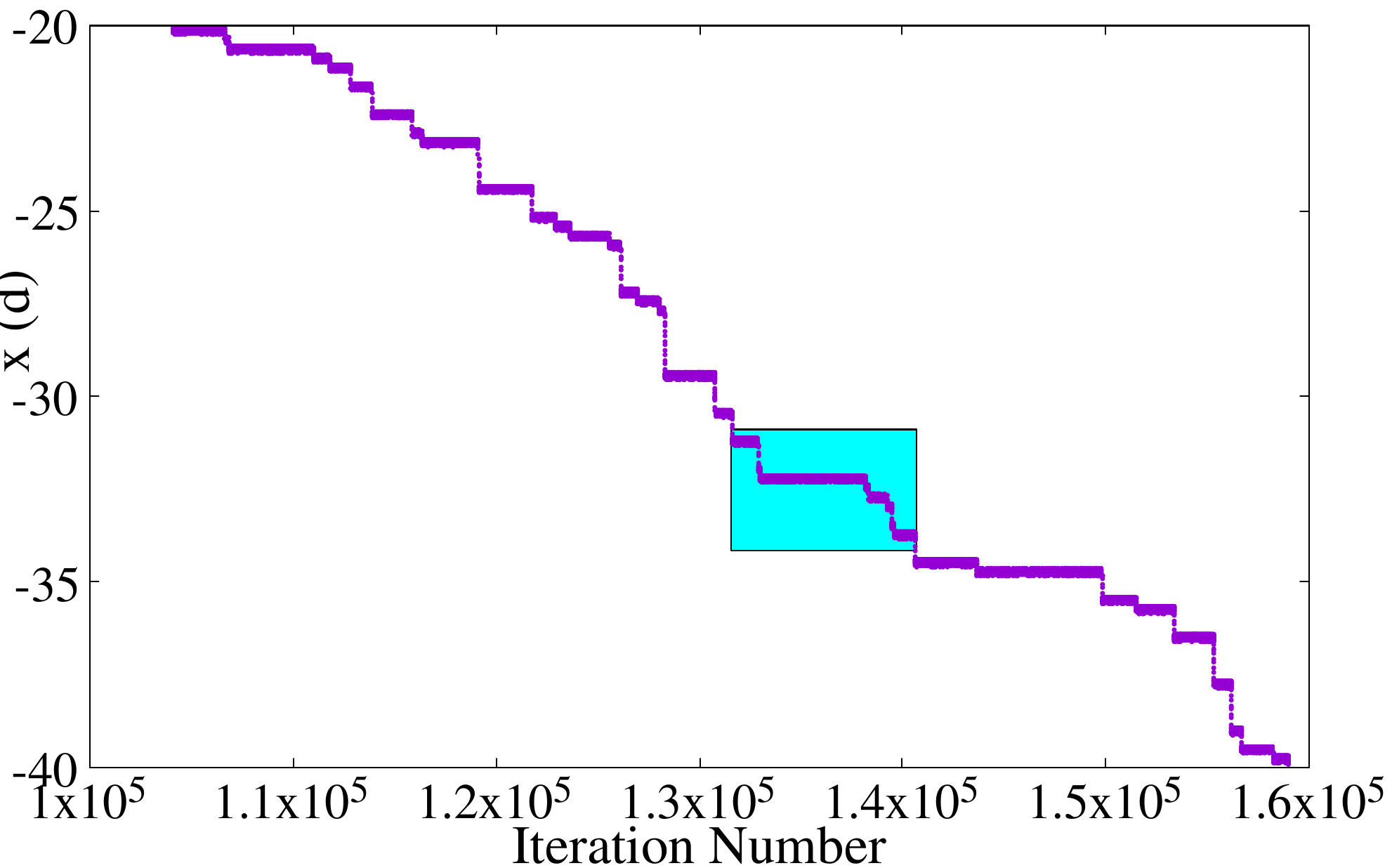}\includegraphics[viewport=5bp 0bp 1063bp 652bp,scale=0.247]{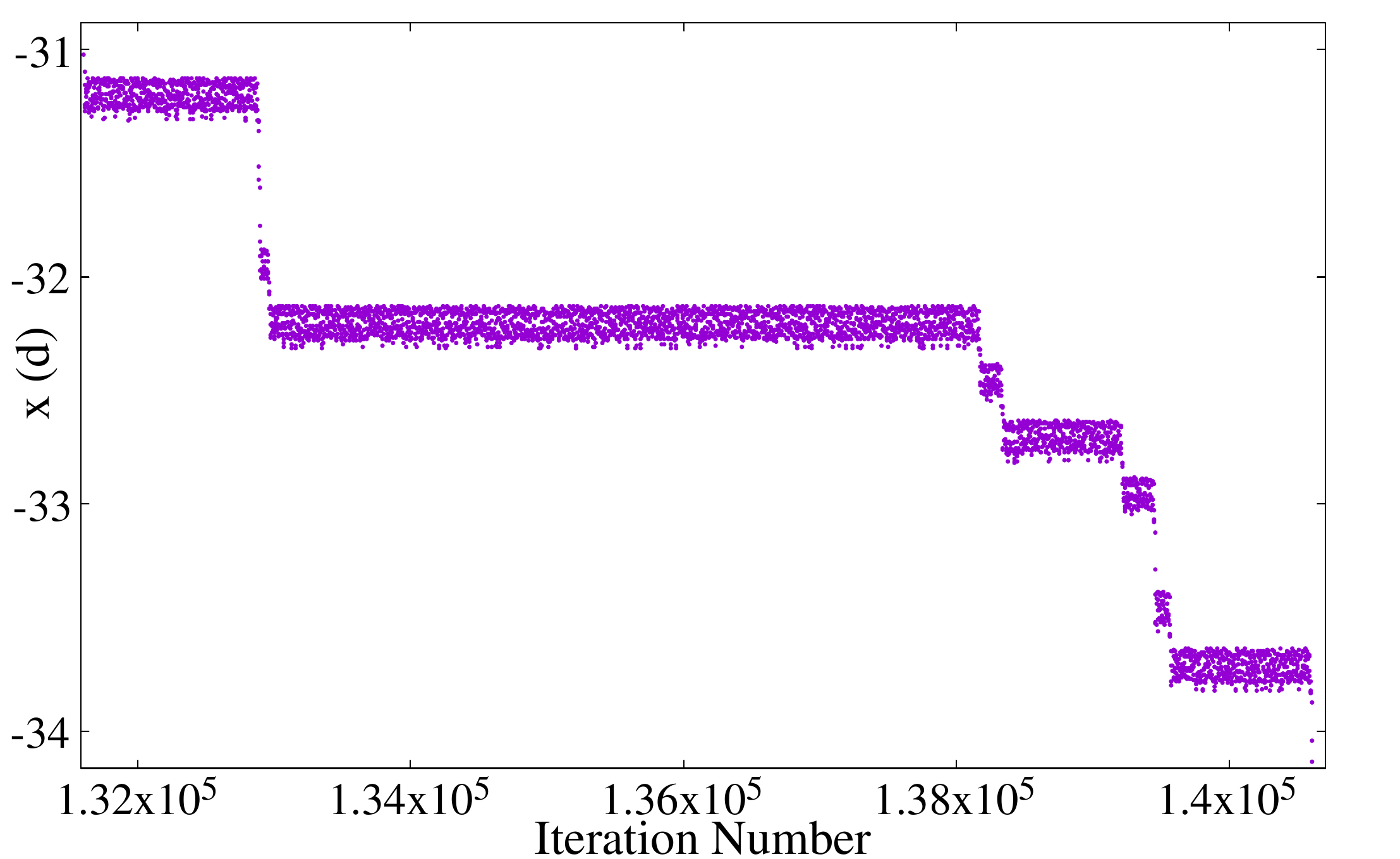}
\caption{Particle trajectory over many iterations for the fixed value of $h=0.3346$.
Other parameters are: $k=25$, the energy cutoff parameter $\gamma_{c}=10^{3}$,
introduced after each intershock space traversing as $1/\gamma\protect\mapsto$1/$\gamma+1/\gamma_{c}$
(see text). The right panel shows a zoom into the rectangle area on
the left.\label{fig:EscapeSC}}
\end{figure*}

We could have left out most of the above as it merely repeats the
well-known properties of quadratic maps, such as the logistic map,
$x\mapsto rx\left(1-x\right)$. Here $r$ is the control parameter,
similar to the shock modulation parameter $h.$ Although our iterated
map is considerably more complex, a quadratic map near an FP can approximate it. Hence, our FP splittings (period-doubling
bifurcations) reproduce those observed in the logistic map to the finest details.
Feigenbaum first discovered them in his pioneering studies.
The morphological equivalence is unmistakably recognizable from comparing
logistic map diagrams with those shown in Fig.\ref{fig:PerDoublBif}.
The bifurcation $h-$values, $h=h_{n}$, in our map can be related
to the respective values of $r_{n}$. However, it is easier to demonstrate
the equivalence of the routes to chaos in these two models by using
general laws that the sequence $h_{n}$ must obey, again found by
Feigenbaum. In particular, the Feigenbaum constant,
\[
\delta\equiv\lim_{n\to\infty}\frac{h_{n}-h_{n-1}}{h_{n+1}-h_{n}}\approx4.67,
\]
defined here for $h$ instead of $r$ in the logistic map, must be
the same. Taking the required $h-$ values from the plots in Fig.\ref{fig:PerDoublBif}
even for a relatively low value of $n=4$, we find $\delta\approx4.57$.
This value is close enough to Feigenbaum's universal constant,
given the low value of $n$.

\begin{figure}%
\includegraphics[scale=0.25]{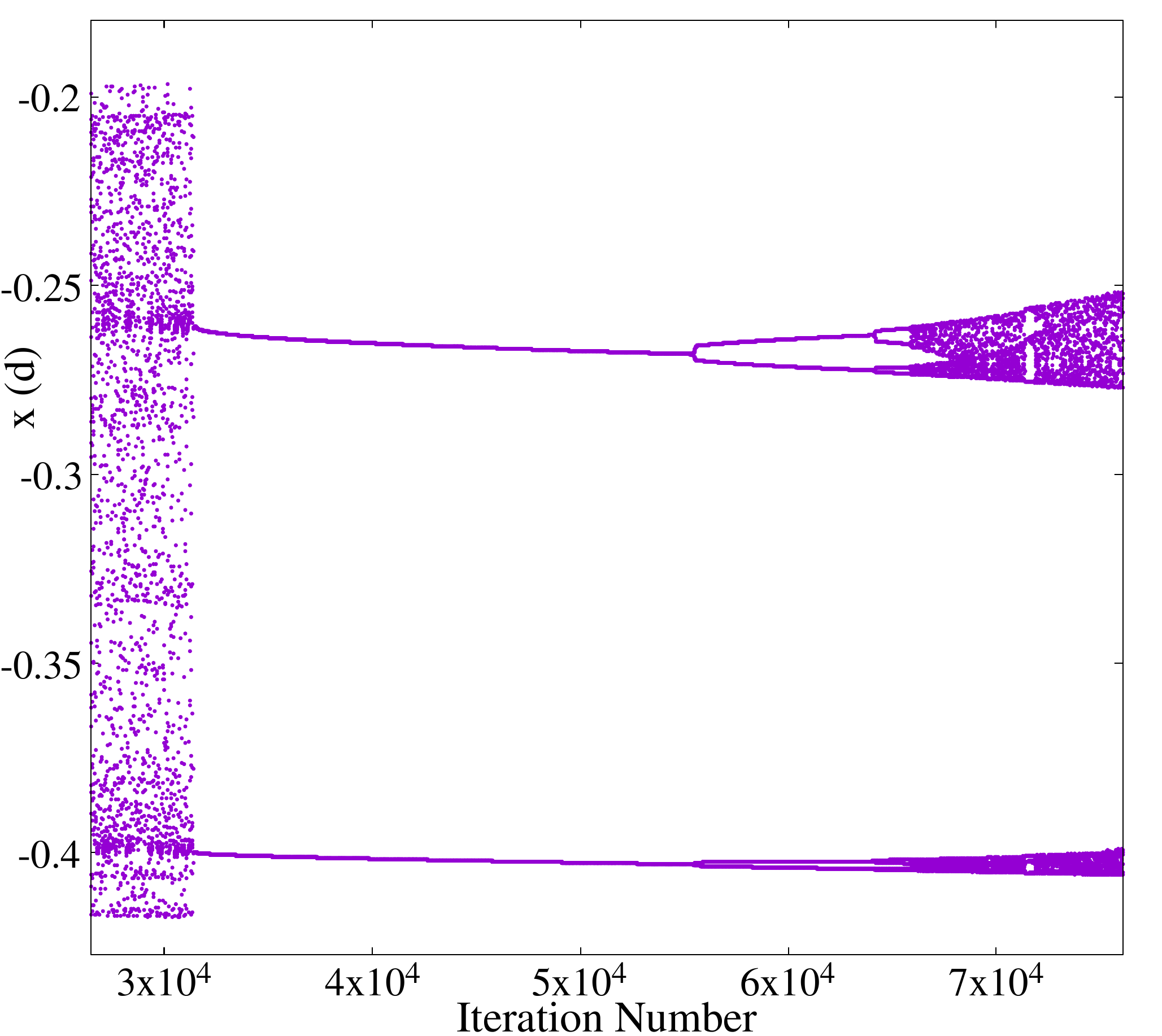}

\caption{Particle trajectory for the shock corrugation parameter $h$ slowly
raising between $0.4615<h<0.4873$, while the iterations proceed.  Beyond  $h\approx 0.4873$, particles escape
in a way similar to that shown in Fig.\ref{fig:EscapeSC}. \label{fig:TrapToEscapeBif}}
\end{figure}%

One may observe in Fig.\ref{fig:PerDoublBif} that the range of stochastic
motion starts to broaden toward smaller $x$ when $h$ exceeds $h\approx0.31$.
Below that value, the $x-$ range is shorter than the gap
between the shocks so that we can rule out particle escape for $h<0.31$.
As $h$ increases, the diffusive particle propagation within a limited
$x-$ interval changes to mixed ballistic-diffusive dynamics that
ultimately result in particle escape. Nevertheless, it does not necessarily
terminate the acceleration but almost certainly affects the spectrum
of accelerated particles. To illustrate such behavior, we iterate
the map at fixed $h\approx$0.33, as shown in Fig.\ref{fig:EscapeSC}.
In the long run, particles propagate ballistically. The average displacement
along the shock front grows linearly with time (number of iterations).
However, zooming into a short piece of particle orbit shows a ``staircase''
transport, which is indicative of L\'evy flights: long-lasting stochastic
traps are interspersed with quick jumps to the next trap, e.g., \cite{Zaslavsky2002}.

This type of particle transport along the shock front continues to
higher $h-$ values. Knowing that the inverse bifurcations to a regular
motion are possible, we have scanned the parameter space in $h$ beyond
its value used in Fig.\ref{fig:EscapeSC}. Indeed, at $h\approx0.465$, a
narrow (in $x$) stochastic trap collapses to a period-two orbit,
as seen in Fig.\ref{fig:TrapToEscapeBif}. This state persists
only while $h$ increases up to $h+\Delta h$, with $\Delta h\sim0.01$,
after which an ordinary sequence of period-doubling bifurcations follows.
After reaching a chaotic state when $h$ is increased by another bit
$\sim0.01,$ particle transport returns to a L\'evy flight type,
as described earlier and illustrated in Fig.\ref{fig:EscapeSC}. 

\begin{figure}%
\includegraphics[scale=0.28]{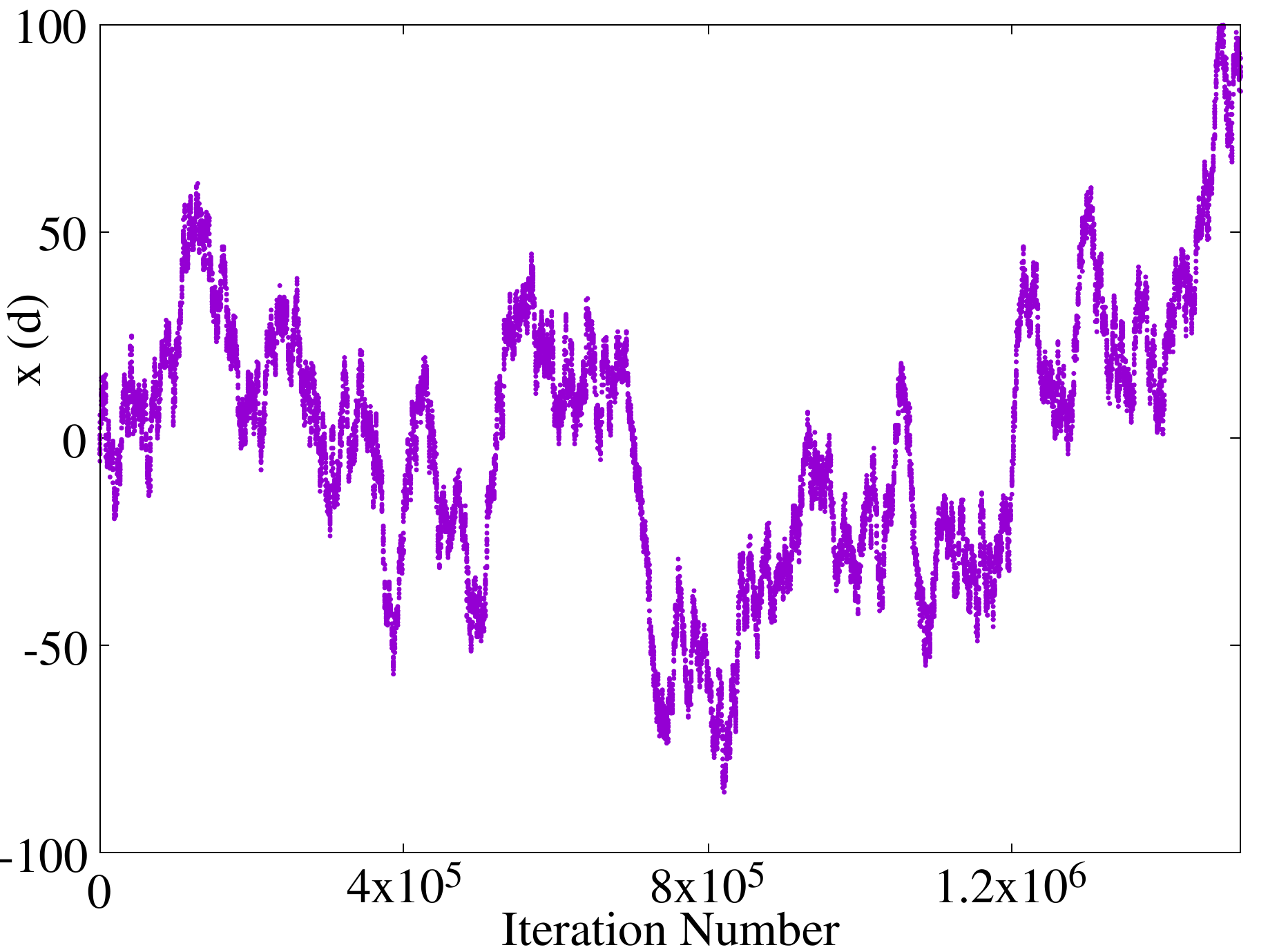}
\caption{Particle transport in the case of both shock surfaces perturbed
according to eq.(\ref{eq:Corrug-d}) with the same amplitude $h=0.45$,
wave-number $k=25$ but shifted in phase by $\pi/2$.\label{fig:DiffWithTwoShockPert}}
\end{figure}%

Up to now, we have deliberately considered the most simple perturbation
of the shock surface. It randomizes particle dynamics and controls
their escape. This approach allowed us to identify the period-doubling
route to the chaos, which is based on the loss of stability of FPs of the
map within a range of control parameters. Since our system has some
similarities to billiards, especially for non-relativistic shocks, this
analogy suggests that the FP stability in corrugated shocks is associated
with the concave (as seen from the intershock space) parts of shock
surfaces. As in the billiards, it has a focusing effect on particle
orbits. By contrast, the convex parts destabilize FPs. We may surmise
that by perturbing both shocks instead of one and by shifting the
phase between the perturbations while holding the wave numbers equal,
we may intermingle the focusing and defocusing portions of the two
shocks. One can expect stronger chaos that might result in the
diffusive rather than L\'evy transport. We show the result of this
numerical experiment in Fig.\ref{fig:DiffWithTwoShockPert} for the
phase shift between the rippling of two shocks by $\pi/2.$ We can
roughly attribute this particular choice of the shocked layer perturbation
to a bending mode instability of thin shocked layers (see, e.g., \cite{Sotnikov2020}
and references therein). 

\begin{figure*}
\includegraphics[viewport=0bp 0bp 882bp 546bp,scale=0.285,viewport=0bp 11bp 749bp 556bp]{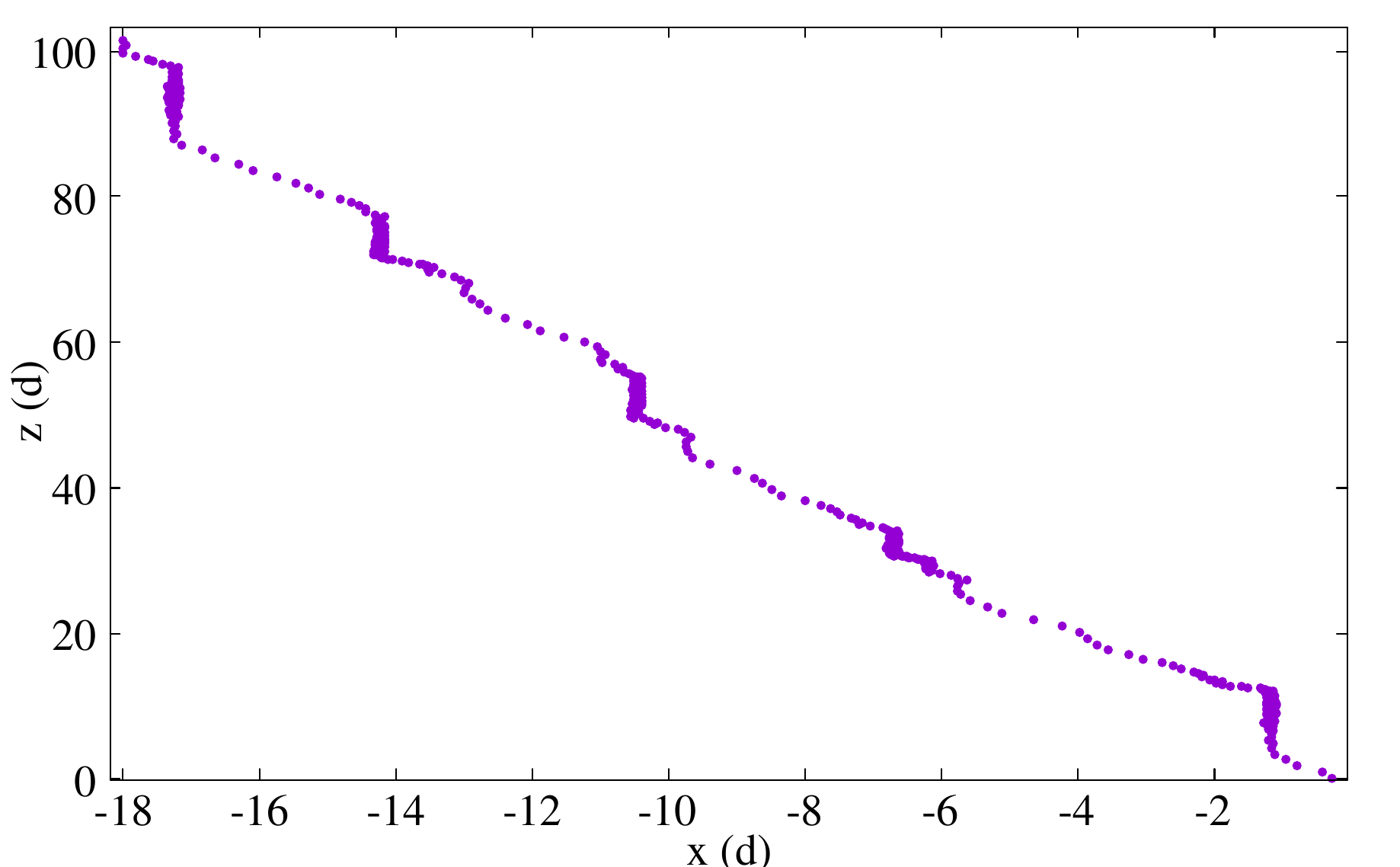}\includegraphics[viewport=-100bp 15bp 882bp 516bp,scale=0.305]{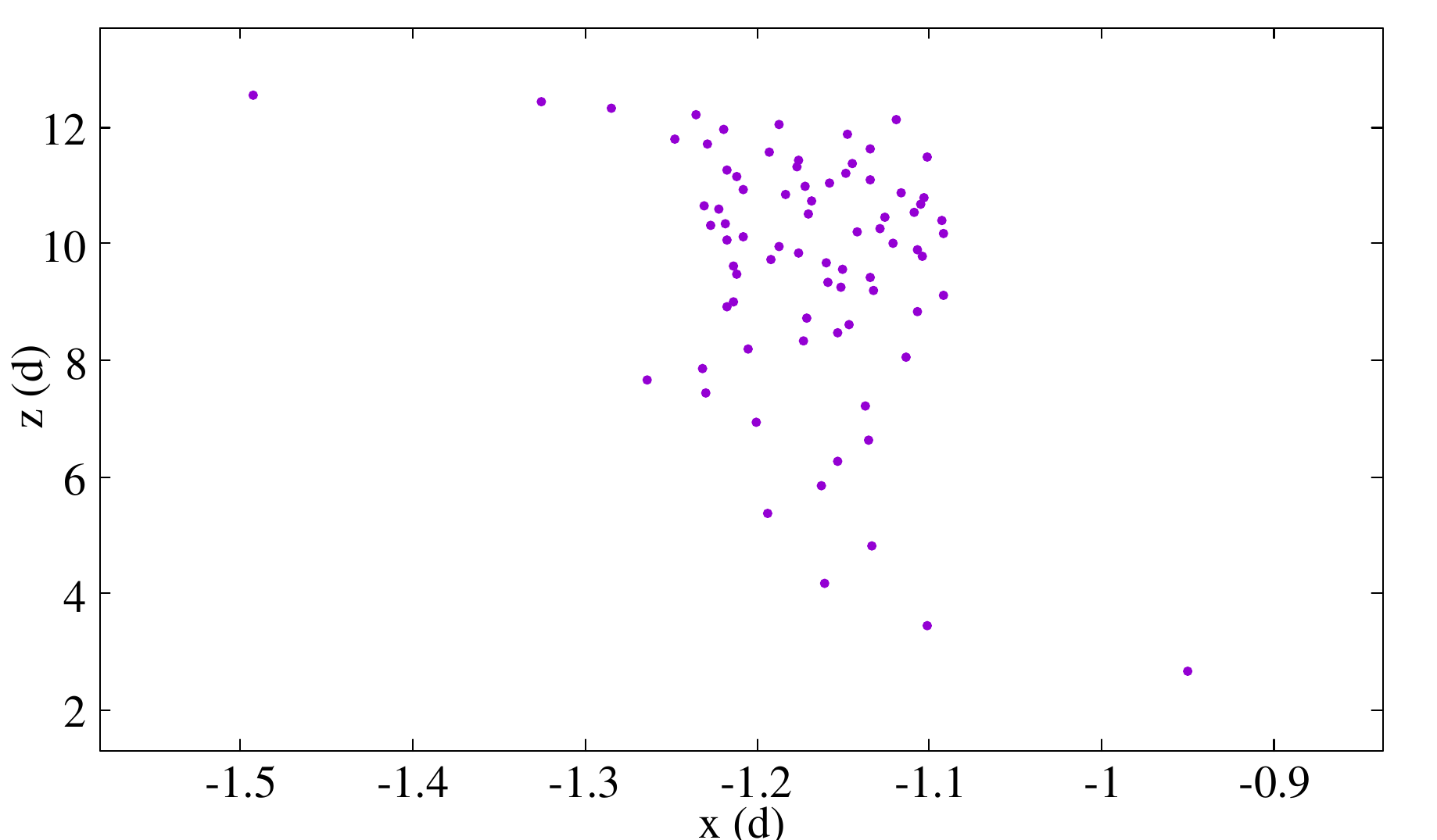}
\caption{Iterated map of particle crossings of one of the shocks shown on its
plane. The angle between the upstream fields is $17.5^{\circ},$ the
shock velocities are $u_{1}=u_{2}=0.3$, and the shock modulation
parameter of one of the shocks is $h=0.3345$, while the other shock
is planar. The right panel shows a zoom into one of the
localized point clusters (see text). \label{fig:DriftOneShockPert}}
\end{figure*}

\subsection{Nonparallel Fields}\label{subsec:Nonparallel-Fields}
Our final example of the double-shock acceleration illustrates the
effect of nonparallel magnetic fields upstream. If we apply the geometrical
consideration of parallel fields given in Secs.\ref{sec:IteratedMap}
and \ref{sec:Properties-of-Map} to antiparallel ones, we find that
the horizontal particle displacements cannot be compensated after
two consecutive collisions with the shocks. This compensation was
necessary to reach an FP and owed to the same sense of rotation in
both upstream media. By contrast, particles will now drift along their
fronts and escape, thus rendering this field geometry unfavorable
for acceleration. We, therefore, consider the case in which the angles
between the fields are less than $90^{\circ}$. In this case, particles
still drift toward a direction in between the fields but presumably slower
than when the angle is larger than $90^{\circ}$. Moreover, by adding
shock corrugation to the nonparallel field configuration, we will explore
the possibility of slowing this drift down.

\begin{figure*}
\includegraphics[viewport=30bp 0bp 1081bp 780bp,scale=0.16]{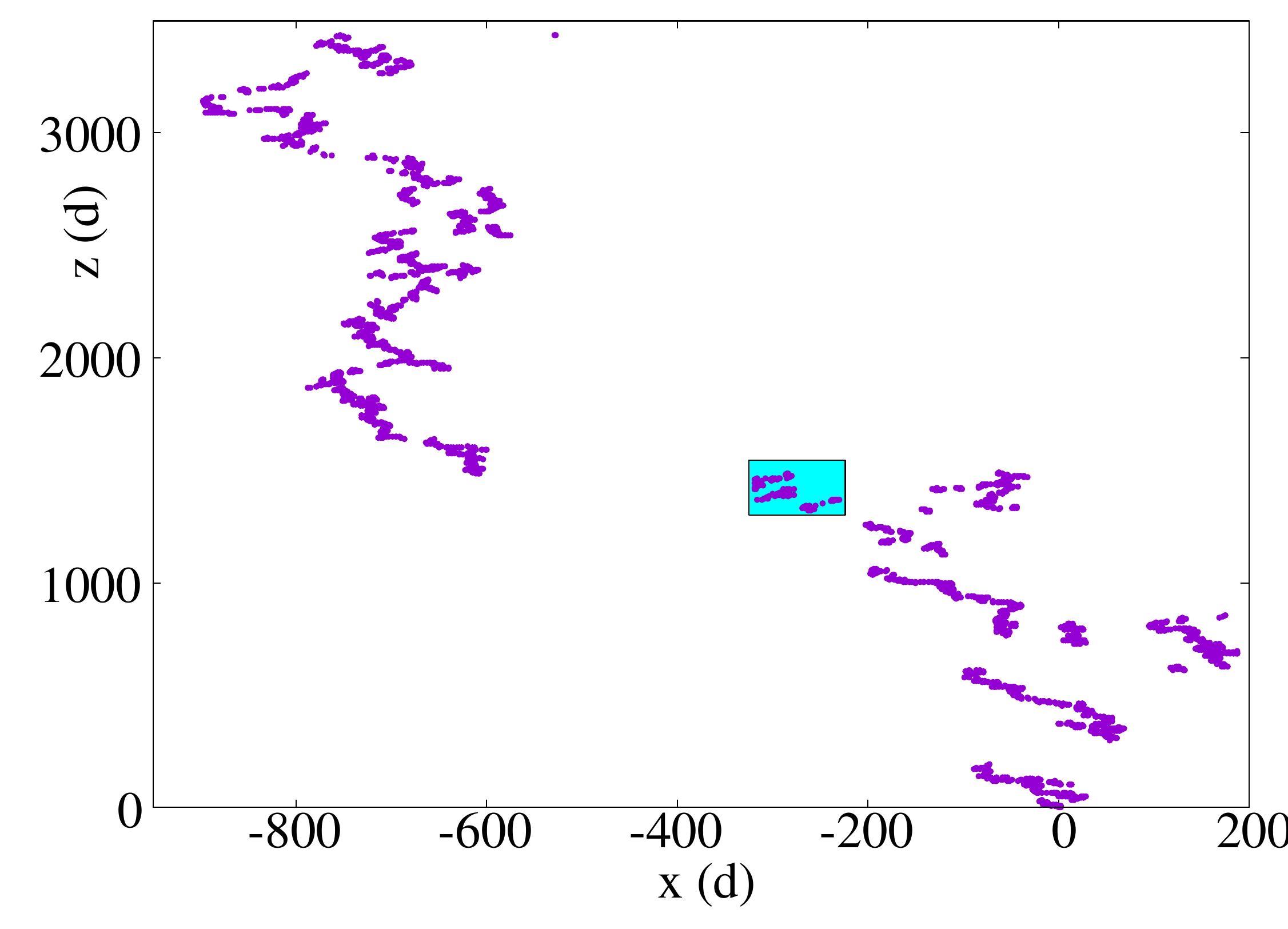}\includegraphics[viewport=30bp 19bp 920bp 671bp,scale=0.193]{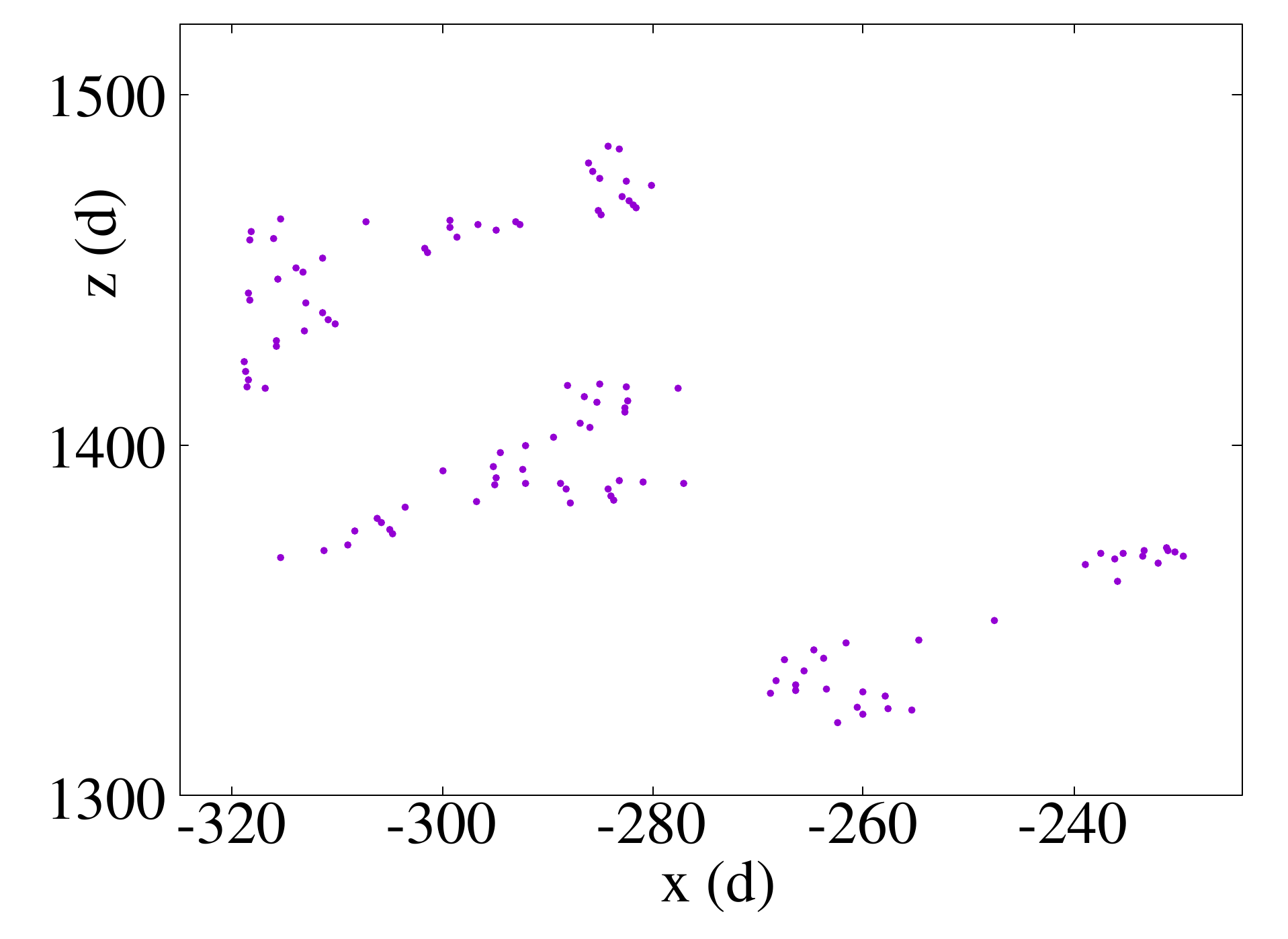}\includegraphics[scale=0.188]{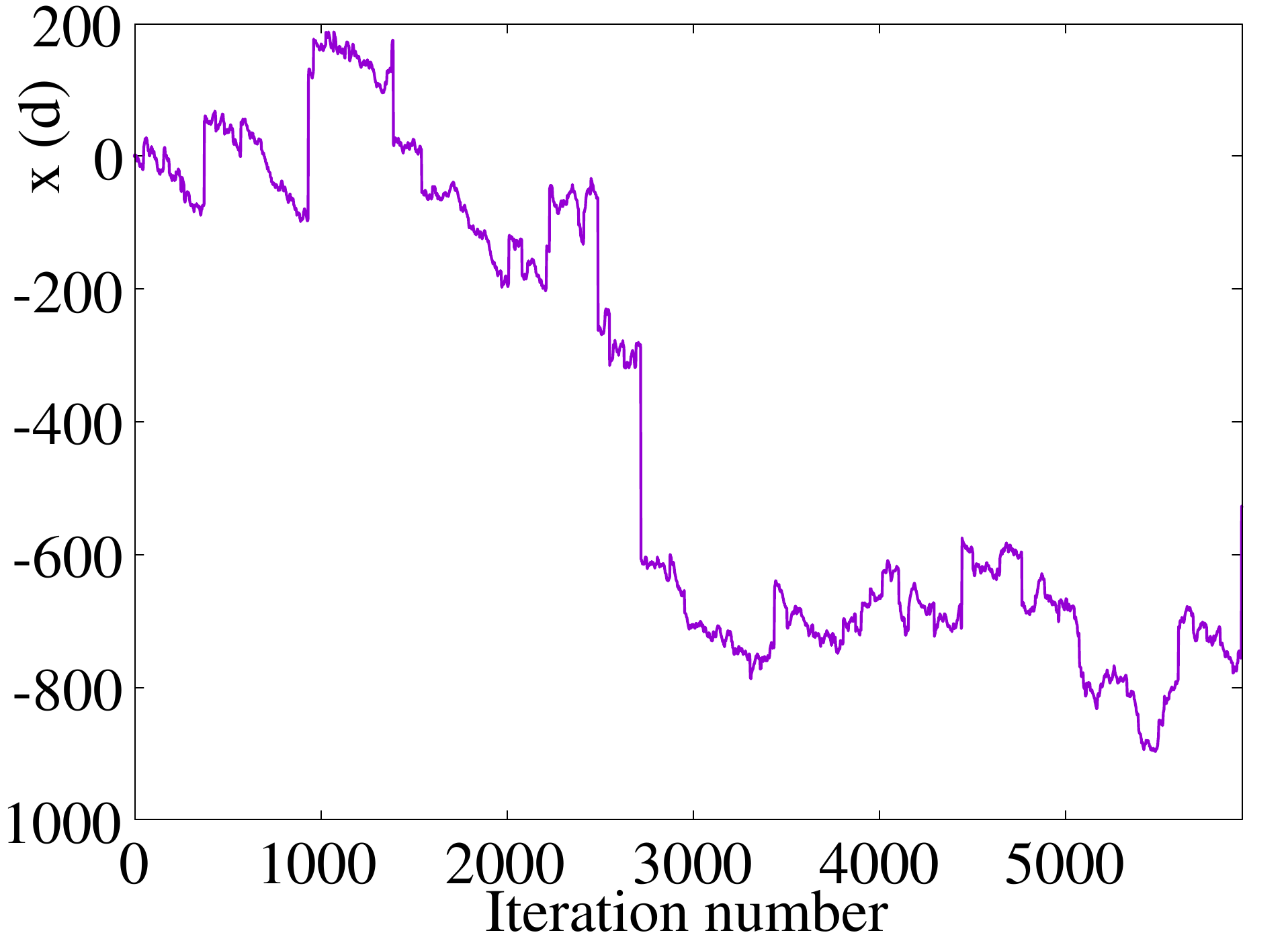}
\caption{Same as in Fig.\ref{fig:DriftOneShockPert} except the shocks have
different speeds, $u_{1}=0.7$ and $u_{2}=0.9$ and the angle between
the upstream fields is $\vartheta=40.3^{\circ}$. Both shocks are modulated
with the same amplitude, $h=7.5$, wave number $k=75$, and no phase
shift between modulations. The middle panel shows a zoom into a typical
cluster of points in the left panel. The right panel shows the $x-$
projection of the particle trajectory, depending on the number of
iterations (the equivalent of time). The total number of iterations shown
in the left and right panels is $\simeq6\cdot10^{3}$.\label{fig:LevyEscape2Dchaotic}}
\end{figure*}

Shown in Fig.\ref{fig:DriftOneShockPert} is the drift in crossed
fields with a small amplitude corrugation added to one of the two
shocks. The particle trajectory is visualized on the $x-z$ plane
of the shocks ($x$ is directed along the corrugation wave vector,
$z$ - along the magnetic field). Although the particle
drifts across the shock plane, it only moves by a $\sim100$ of intershock
distances after $10^{3}$ iterations. More importantly, the drift proceeds
in a ``wait-flight'' fashion, whereby the particle is quasi-periodically
\emph{localized} to small spots for many chaotic collisions before
it strides to sample another cluster of points. These chaotic clusters' relatively regular,
rectangular shape is associated with 
one-shock corrugation. Indeed, perturbed surfaces of both shocks
(still only in $x$ direction) make the particle transport more random. 

We illustrate a more chaotic transport in Fig.\ref{fig:LevyEscape2Dchaotic}.
To show the self-similarity of a chaotic orbit attractor, we present
a blow-up of one of its clusters on the $x,z$ plane. It is seen that
the cluster consists of smaller clusters that, in turn, show some
further clustering until the number of points becomes insufficient
to follow through the self-similarity. To highlight the L\'evy flight
events, we show the same trajectory in a time-dependent representation
by plotting the $x$- coordinate as a function of the iteration number.
The latter can be regarded as a proxy of time. Although the long-time
particle propagation is significant; its main contribution comes
from the long strides. Between them, the particle bounces between
the shocks, remaining well confined to a chaotic cluster. During this
time, it may reach the energy cutoff, thus making the escape process
not critical for the maximum energy.

\section{Summary and Discussion}\label{sec:Discussion}

In this paper, we have studied a particle acceleration mechanism that
operates in a double-shock system formed in colliding plasma flows.
The acceleration is powered by upstream flows of the shocks that bound
the shocked plasma layer. High energy particles bounce between these
flows until they are convected with the shocked plasma or otherwise
escape the accelerator. Our main findings are as follows:
\begin{enumerate}
\item The acceleration mechanism requires seed particles, pre-energized
before entering the intershock space. Their gyroradius needs to be
$r_{\text{g}}\gg\sqrt{ld}$, where $l$ is the principal turbulence
scale in the shocked layer, and $d$ is its thickness (see, however, Appendix \ref{sec:Appendix-A}). Under most favorable
acceleration conditions, these particles reach equilibrium and
bounce between fixed points on the shock surfaces for a long time.
It happens, e.g., when the magnetic fields upstream are parallel, and 
the shocks propagate at the same speed. 
\item If these shocks are non-relativistic, $u\ll c$, the energy gain factor
per collision with each shock is $1+2u/c.$ For $u\to c$, it increases
to $\approx3.73$ for a particle equilibrium ingress angle, $\alpha=\alpha_{\text{eq}}\approx\negthinspace\negthinspace35.3^{\circ}$,
and reaches a higher value, $\approx4.15$, for an optimum angle,
$\alpha_{\text{opt}}\simeq\alpha_{\text{eq}}$, depending on $u$.
\item If the shock configuration is not ideal, particles may drift along
the shock front and ultimately escape. We have considered the following
three mechanisms of escape:
\begin{enumerate}
\item \emph{Unequal shock speeds}. This effect is potentially significant for non-relativistic
shocks, and it naturally increases with the shock speed difference,
$\Delta u=u_{1}-u_{2}$. It is much less important for relativistic
shocks, in part because of their high energy gain per cycle.
\item \emph{Shock corrugation}. For a sinusoidal corrugation of at least one 
of the shocks, the particle bouncing between them becomes chaotic. For low-amplitude
corrugations, though, particles do not propagate along the shock surface
and, therefore, do not escape. Stronger corrugations trigger quasi-ballistic
particle flights and eventual escape. However, the flights are occasionally
interrupted when particles get trapped into chaotic local attractors
and keep bouncing there, thus gaining high energy before jumping
to another chaotic attractor.
\item \emph{Nonparallel magnetic fields upstream}. This field configuration results
in a systematic particle's drift and escape, especially if the angle
between the fields is large, $\vartheta\sim1$ or even more so if the
fields are close to being antiparallel, $\vartheta\approx\pi$. However,
an addition of a significant shock corrugation strongly diminishes
the drift. Its suppression is similar to that described in (b): long
L\'evy flights are interspersed by particle trapping in localized
regions of chaotic motion.
\end{enumerate}
\end{enumerate}
We conclude that the most efficient and sustainable acceleration
occurs between relativistic shocks of the same speed with parallel
magnetic fields upstream. The field strengths do not have to be equal.
Although these are still highly idealized conditions, they give a
good starting point for assessing the energy spectrum of accelerated
particles. For most Fermi acceleration mechanisms, the energy spectral
index can be calculated as $q=1+\tau_{\text{acc}}/\tau_{\text{esc}}$,
where $\tau_{\text{acc}}$ and $\tau_{\text{esc}}$ stand for the
acceleration and escape times. The former is shorter than the particle
travel time between shocks multiplied by $c/u$, while the escape
time is infinite under the idealized conditions so that we obtain
$q=1$ in this case. 

The extreme spectral index, $q=1$, may be extended to a broader class
of relativistic shocks, based on the results of Sec.\ref{sec:Particle-Energy-Loss},
whenever $\tau_{\text{acc}}\ll\tau_{\text{esc}}$. By contrast, the
spectrum may substantially soften for non-relativistic shocks, depending
on their deviation from the ``ideal'' acceleration conditions. In
any event, we are dealing with a \emph{reacceleration} mechanism, where the end spectrum may depend on the seed particle spectrum
as much as on the properties of the accelerator \cite{1994ApJ...431..705S},
which was argued by \cite{MalkMosk2022} to be supported by recent
precise observations \cite{Aguilar2021}. Besides, the particle escape
time is sensitive to several parameters that require modeling.
While the situation with the index is much clearer for relativistic
shocks, the uncertainty about the maximum energy is significant and
should also be addressed on the case by case basis. Nevertheless, we may
draw some conclusions regarding the acceleration parameters based
on the results obtained in this paper.

Since the acceleration rate for a pair of relativistic shocks is very
high,  the maximum energy will likely be limited
by energy losses rather than particle escape, particularly in extended systems. In this case, the accumulation
of particles toward the cut-off, determined by the energy losses,
will almost certainly occur. This observation returns us to the
problem of hard spectra with $q<1$, e.g., \cite{Globus2015},
already mentioned in the Introduction. Due to the likely particle
pile-up toward the cut-off, such a hard spectrum can form. The precise
spectral shape will be controlled by the triad of acceleration-, escape-,
and energy-loss rate. The escape rate is most challenging to quantify
because it is very sensitive to shock corrugations, as our results
indicate. This is a difficult subject in shock studies, especially
in conjunction with particle acceleration \cite{MondDrury98}
and relativistic shocks \cite{2016ApJ...827...44L,Demidem2018}. 

One suggestion that comes from the proposed acceleration mechanism
is to look for possible high-energy counterparts of periodic X-ray
outbursts observed in binary stars~\cite{Corcoran1997,2009ApJ...698L.142T,Pollock2021}.
High-energy particles needed for the double-shock acceleration may
be seeded from the ordinary diffusive shock acceleration, whose efficiency
may be correlated with the outbursts. We then invoke the requirements
for efficient particle acceleration. The most critical requirement
is the collinearity of the magnetic fields upstream of the shocks.
In a binary system, this condition can be met only
at specific phases of orbital and stellar rotation. Given that
the magnetic and rotation axes are unlikely to coincide, the Parker
spiral fields upstream of the colliding shocks of the stars near their centerline
is almost certainly highly variable due to the stellar and orbital rotations. This area
is the ``hot spot'' of the particle acceleration mechanism.
The field angle $\vartheta$ will then change in time quasi-periodically,
bearing on the star rotation periods, orbital period, and wind variability. However, $\vartheta$  will
occasionally approach $\vartheta=0$ when the
efficient acceleration should burst.

\acknowledgments

MM acknowledges support from the National Science Foundation under grant No. AST-2109103.
ML acknowledges support from the ANR (UnRIP project, Grant No.~ANR-20-CE30-0030).

\appendix

\section{Particle Transport between Shocks}\label{sec:Appendix-A} 
Below, we investigate
the propagation of energetic test particles across a plasma compressed
between two shocks. We have assumed throughout the paper that the
particle energy is high enough to make the propagation rectilinear.
Here we take a broader look to ascertain which gap and how a particle
can cross, not necessarily in a straight-line fashion.  Alternatively, it
returns back to the shock it came from. The straight-line propagation
criterion, estimated in Sec.\ref{sec:Double-Shock-Acceleration-Model}, will then be justified
and refined. In addition, we will demonstrate that particles will
start gaining energy by multiple interactions with one or both shocks,
after which they gain enough energy to propagate rectilinearly.

We assume the shocked
plasma outflows along the contact discontinuity between the shocked
gases. The velocity component normal to the shock fronts is neglected in treating
the transport of energetic particles. Their Larmor radius exceeds
the transition zone where the flow behind each shock turns its direction
from normal to tangential; the Larmor radius can still be comparable with
the gap between
the shocks, $d$. We also assume that particles enter the intershock space and
propagate towards the opposite shock in the magnetic field.
It consists of two components: a compressed (by the shock compression
ratio) regular upstream field. It remains parallel to both shocks, 
and its strength is $\boldsymbol{B}_{0}=B_{0}\boldsymbol{e}_{z}$.
The second component is turbulent, $\delta B\gg B_{0}$, that has
a short correlatioin length, $l\ll r_{g},$ where $r_{g}=cp/e\delta B$.

Particle propagation in
such a short-scale turbulent field has been considered in detail,
including numerical simulations evolving many individual
trajectories in \citep{casse2001transport,Plotnikov_2011}. For the
above-described purpose, a simpler approach can be employed. Namely,
we solve a kinetic equation averaged over $x,z$ -variables along
the shock fronts.
After transforming to polar coordinates in velocity space, $\left(v,\phi,\vartheta\right)$,
with $v_{z}$ component along the polar axis and unperturbed field
$\boldsymbol{B}_{0}$, $v_{z}=v\cos\vartheta$, $v_{x}=v\sin$$\vartheta\sin\phi$,
and $v_{y}=v\sin\vartheta\cos\phi$, the kinetic equation can be written
as

\begin{equation}
\begin{aligned}\frac{\partial f}{\partial t}+c\sin\vartheta\cos\phi\frac{\partial f}{\partial y}+\Omega\frac{\partial f}{\partial\phi}\\
=\nu_{\text{s}}\left(\frac{1}{\sin^{2}\vartheta}\frac{\partial^{2}f}{\partial\phi^{2}}+\frac{1}{\sin\vartheta}\frac{\partial}{\partial\vartheta}\sin\vartheta\frac{\partial f}{\partial\vartheta}\right),
\end{aligned}
\label{eq:KinEqInit}
\end{equation}
where the gyro-frequency
is measured by the unperturbed field $B_{0}$, $\Omega=p/eB_{0},$
$\nu_{\text{s}}$=$cl/r_{g}^{2}$ is the angular diffusion coefficient
(see \citep{Plotnikov_2011} for its dependence on the type and strength
of the turbulent field component). We assume that one of the shocks
is at $y=0$. After averaging eq.(\ref{eq:KinEqInit}) over the pitch-angle
$\vartheta$, we obtain: 
\begin{equation}
\frac{\partial F}{\partial t}+c\cos\phi\frac{\partial}{\partial y}F\overline{\sin\vartheta}+\Omega\frac{\partial F}{\partial\phi}=\nu_{\text{s}}\frac{\partial^{2}}{\partial\phi^{2}}F\overline{\sin^{-2}\vartheta}\label{eq:KinEqAver}
\end{equation}
Here 
\[
F\equiv\int_{0}^{\pi}f\sin\vartheta d\vartheta,
\]
and

\[
\overline{\sin\vartheta}\equiv\frac{1}{F}\int_{0}^{\pi}f\sin^{2}\vartheta d\vartheta,\,\,\,\,\,\,\,\,\,\,\,\,\,\,\,\overline{\sin^{-2}\vartheta}\equiv\frac{1}{F}\int_{0}^{\pi}f\frac{d\vartheta}{\sin\vartheta}
\]
It is reasonable to assume that $f\to0$ for $\sin\vartheta\to0$
since such particles escape the intershock space, thus making the
last integral converge. The last two quantities can be estimated differently
for relativistic and nonrelativistic shocks. According to Sec.\ref{sec:Properties-of-Map},
relativistic shocks quickly bring the mapping to the state with $\sin\vartheta\approx1$,
so we can set $\overline{\sin\vartheta}\approx\overline{\sin^{-2}\vartheta}\approx1$,
which is accurate with a possible exception for a few first iterations.
For nonrelativistic shocks, the pitch angle does not change significantly
between acceleration cycles and we can set $\vartheta=\vartheta_{0}={\rm const.}$,
as we are essentially interested in a single particle behavior (Green's 
function) for mapping purposes. By introducing dimensionless variables
according to $\Omega t\to t$, $y/r_{g}^{*}\to y$, where $r_{g}^{*}\equiv c\overline{\sin\vartheta}/\Omega\sim r_{g0}$
(with $r_{g0}$ measured by $B_{0})$ , eq.$\left(\ref{eq:KinEqAver}\right)$
rewrites as follows:
\begin{equation}
\frac{\partial F}{\partial t}+\cos\phi\frac{\partial F}{\partial y}+\frac{\partial F}{\partial\phi}=D\frac{\partial^{2}F}{\partial\phi^{2}}\label{eq:KinEqND}
\end{equation}
We have introduced the dimensionless scattering frequency $D$=$\nu_{\text{s}}\overline{\sin^{-2}\vartheta}/\Omega$.
We note that $D$ is the only parameter in eq.(\ref{eq:KinEqND})
and it is worthwhile to assess its range. It can be written as $D\sim\nu_{s}/\Omega\sim\left(l/r_{g0}\right)\left(\delta B/B_{0}\right)^{2}$,
thus being a product of small and large parameters. We thus assume
$D\sim1$, but start with simple limiting cases of ballistic and diffusive
transport regimes.
For ballistic propagation,
$Dt\ll1$, we have a general solution, 
\[
F\left(t,y,\phi\right)=\mathcal{F}\left(\phi-t,y-\sin\phi\right)
\]
where $\mathcal{F}$ is an arbitrary function to be obtained from
initial and boundary conditions. In general, the latter should be
specified at a curve on $\left(y,\phi\right)$ -plane not coinciding
with any of characteristics (particle orbits) $y-\sin\phi=const$.
In this Appendix, we focus on an initial condition given at the shock
at $y=0$ and follow the propagation of particles entering the intershock
gap at angle $\phi_{0}$ to its normal. A normalized solution for
$F$ can be written for this case as 
\begin{equation}
F=\delta\left(\phi-\phi_{0}-t\right)\delta\left(y+\sin\phi_{0}-\sin\phi\right)\label{eq:BallistSol}
\end{equation}
which is, of course, just a Larmor rotation. A particle can reach
the opposite shock at $y=d/r_{g}^{*}$, if $1-\sin\phi_{0}>d/r_{g}^{*}$.
The rectilinear propagation regime strictly requires
\begin{equation}
1-\sin\phi_{0}\gg d/r_{g}^{*},\label{eq:RectLinCond}
\end{equation}
which was specified in Sec.\ref{sec:Double-Shock-Acceleration-Model}. The above condition
can be weakened to $1-\sin\phi_{0}>d/r_{g}^{*}$, if a curved particle
trajectory, $y_{0}\left(t\right)=\sin\left(\phi_{0}+t\right)-\sin\phi_{0}$,
is included into the iterated map derived in Sec.\ref{sec:IteratedMap},
instead of the straight-line trajectory. In the opposite case, $1-\sin\phi_{0}<d/r_{g}^{*}$,
particles return to the shock at $y=0$, and enter the next acceleration
cycle. If $d/r_{g}^{*}>2$, all particles will do just that, but after
several cycles, they will gain enough energy to cross the gap and ultimately
enter the rectilinear propagation regime.

In the opposite case,
$Dt\gg1$, by expanding eq.$\left(\ref{eq:KinEqND}\right)$ in $1/D$-
series we arrive at a diffusive equation:
\begin{equation}
\frac{\partial}{\partial t}\left\langle F\right\rangle =\frac{1}{2D}\frac{\partial^{2}}{\partial y^{2}}\left\langle F\right\rangle ,\,\,\,\,\,\,\,\,\,\text{where}\,\,\,\,\,\,\,\,\left\langle \cdot\right\rangle \equiv\frac{1}{2\pi}\int_{0}^{2\pi}\left(\cdot\right)d\phi.\label{eq:DiffLimit}
\end{equation}
This equation can be used to set a limit to a diffusion time between
shocks, $t_{d}\sim\left(d/r_{g}^{*}\right)^{2}D$, that should be
shorter than the convection time along the shock fronts, $L_{\rm sh}/u_{\text{conv}}$,
where $L_{\rm sh}$ is the shock size. However, eq.(\ref{eq:DiffLimit}) is
separated from the ballistic regime obtained earlier by an extended
transdiffusive regime, in which either approximation is inaccurate.
Therefore, the solution of eq.(\ref{eq:DiffLimit}) cannot be used
for extending the condition in eq.(\ref{eq:RectLinCond}) to the case
$D\neq0$. In the transdiffusive regime at $Dt\sim1$, the diffusive
reduction of eq.(\ref{eq:KinEqND}) is acausal (see e.g., \citep{malkov2017exact}
and references therein). This problem is generic for ``standard''
reduction schemes of a full Fokker-Planck- to the diffusive description.
It is particularly manifested in ``loosing'' the particle gyration
in eq.(\ref{eq:BallistSol}) in the transition from eq.(\ref{eq:KinEqND})
to eq.(\ref{eq:DiffLimit}). The particle gyration effect is expressed
in transient oscillations of the perpendicular diffusion coefficients,
as shown in \citep{casse2001transport,Plotnikov_2011} (see also below).

\begin{figure}
\includegraphics[scale=0.6]{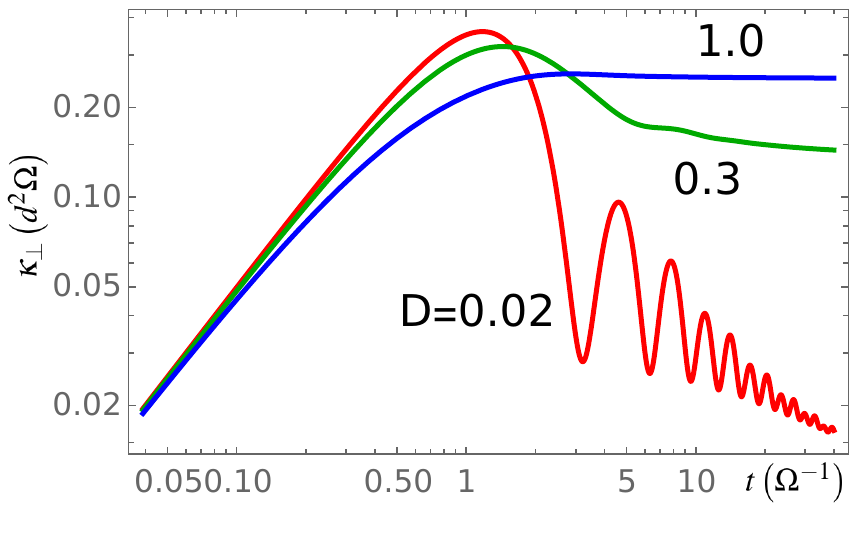}
\caption{Particle diffusion across
the background magnetic field $B_{0}$ as a function of time for different
values of $D$ and $\phi_{0}=0$ (see eq.[{\ref{eq:KappaApA}}])\label{fig:Particle-diffusion-across}}
\end{figure}

We, therefore, return
to eq.(\ref{eq:KinEqND}) to evaluate particle trajectories for an
arbitrary $D$ with no further approximations. First, we
generalize the simple diffusion formula, $\overline{y^{2}}\sim t/D$
that follows from eq.(\ref{eq:DiffLimit}) to an exact expression
for $\overline{y^{2}\left(t\right)}$. For, let us introduce an $n$-th
moment

\[
\overline{y^{n}\left(t\right)}=\int_{-\infty}^{\infty}\left\langle F\right\rangle y^{n}dy
\]
assuming the total number of particles, conserved by eq.(\ref{eq:KinEqND}),
is normalized to unity, $\overline{\left\langle F\right\rangle }\equiv \overline{y^{0}\left(t\right)}=1.$
Assuming further that particles are released at $t=0$ from the plane
$y=0$, from eq.(\ref{eq:KinEqND}) we have
\begin{equation}
\frac{d}{dt}\overline{y^{2}}=2\overline{y\left\langle F\cos\phi\right\rangle }\label{eq:y2bar}
\end{equation}
The correlator $\overline{yv_{y}}$ on the r.h.s. (here $\cos\phi$
is actually the $y-$compononent of particle velocity) characterizes
the particle transport and can be calculated exactly using eq.(\ref{eq:KinEqND})
as follows. First, we introduce a complex function
\begin{equation}
\Theta=\overline{y\left\langle Fe^{i\phi}\right\rangle }\label{eq:ThetaDef}
\end{equation}
for which we have the following equation
\begin{equation}
\frac{\partial\Theta}{\partial t}+D\Theta-i\Theta=\frac{1}{2}\left(W+1\right)\label{eq:Theta}
\end{equation}
 where $W\left(t\right)$ is defined as
\[
W=\overline{\left\langle Fe^{2i\phi}\right\rangle }.
\]
 This function, in turn, can be found from the following equation
\[
\frac{\partial W}{\partial t}+4DW-2iW=0.
\]
Assuming, again, that $F\left(t=0,\phi,y\right)=\delta\left(\phi-\phi_{0}\right)\delta\left(y\right)$,
for $W$ we have
\[
W=e^{2i\left(\phi_{0}+t\right)-4Dt}
\]
From eq.(\ref{eq:Theta}) we obtain
\[
\Theta=\frac{1-e^{it-Dt}}{2\left(D-i\right)}+\frac{e^{\left(i-3D\right)t}-1}{2\left(i-3D\right)}e^{2i\phi_{0}+it-Dt}
\]
After taking $\Re\Theta$ and integrating eq.(\ref{eq:y2bar}), using
eq.(\ref{eq:Theta}) we arrive at the following expression for the
perpendicular diffusion, $\kappa_{\perp}=\overline{y^{2}}/2t$:
\begin{widetext} 
\begin{equation}
{\begin{split}\kappa_{\perp} & =\frac{D}{2\left(D^{2}+1\right)}+\frac{-3\left(D^{3}+D\right)\sin\left(2\phi_{0}\right)+\left(2D^{4}+D^{2}-1\right)\cos\left(2\phi_{0}\right)-8D^{4}+6D^{2}+2}{4\left(D^{2}+1\right)^{2}\left(4D^{2}+1\right)t}\\ & +\frac{e^{-4Dt}\left\{ \left(6D^{2}-1\right)\cos\left[2\left(t+\phi_{0}\right)\right]-5D\sin\left[2\left(t+\phi_{0}\right)\right]\right\} }{4\left(4D^{2}+1\right)\left(9D^{2}+1\right)t}\\ & +\frac{e^{-Dt}\left[4D\left(D^{2}+1\right)\sin\left(t+2\phi_{0}\right)+\left(-3D^{4}-2D^{2}+1\right)\cos\left(t+2\phi_{0}\right)-2\left(9D^{3}+D\right)\sin(t)+\left(9D^{4}-8D^{2}-1\right)\cos(t)\right]}{2\left(D^{2}+1\right)^{2}\left(9D^{2}+1\right)t}\end{split}\label{eq:KappaApA}}
\end{equation}
\end{widetext} {We show $\kappa_{\perp}\left(t\right)$
in Fig.\ref{fig:Particle-diffusion-across} and note that it is consistent
with the perpendicular diffusion coefficient shown in Fig.3 of \citep{Plotnikov_2011}.}

\begin{figure}
\includegraphics[scale=0.57]{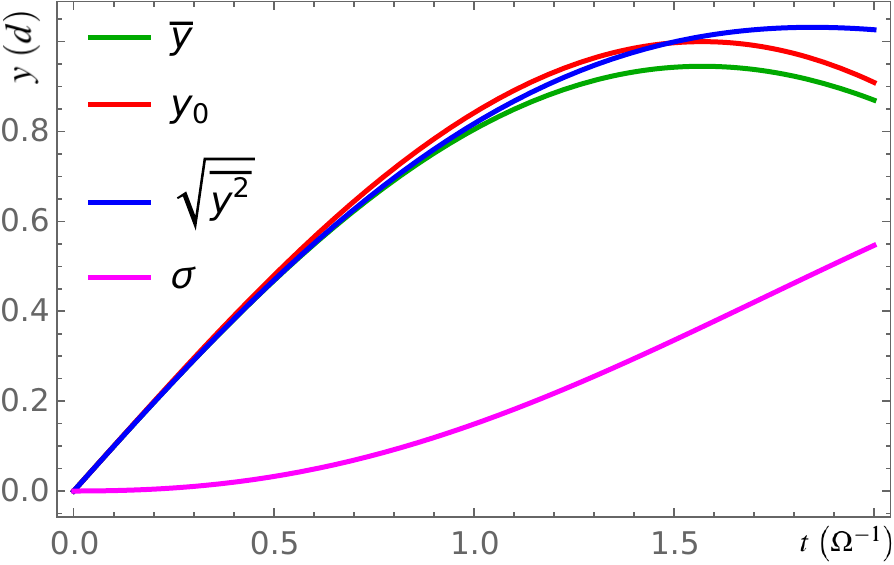}
\includegraphics[scale=0.575]{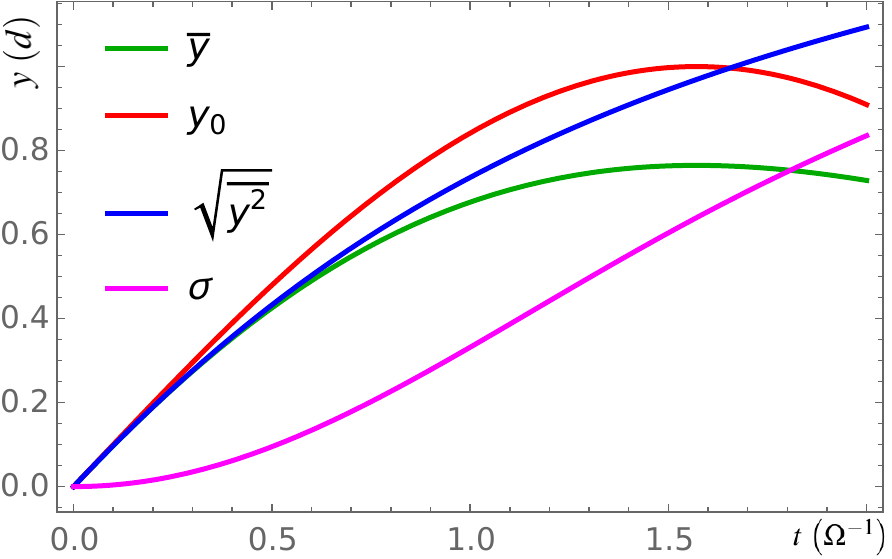}

\caption{Major characteristics
of stochastic particle trajectories compared to the ballistic propagation,
$y_{0}$, shown for $D=0.1$ (upper panel), and $D=0.5$ (lower panel).
The gyro-phase $\phi_{0}=0$ (see text).\label{fig:Major-characteristics-of}}
\end{figure}

To address the question
of the straight-line approximation applicability, let us consider
a mean-square-root deviation of a stochastic particle trajectory $y\left(t\right)$
from a straight line, $y_{0}=y_{\text{sl}}=t\cos\phi_{0}$, or, more
generally, from an arc $y_{0}=y_{\text{arc}}\left(t\right)\equiv\sin\left(\phi_{0}+t\right)-\sin\phi_{0}$,
as the particle moves between $y=0$ and $y=d/r_{g}^{*}$. To obtain
the rms deviation of the stochastic trajectory from the ballistic one,
$\sigma=\sqrt{\overline{\left[y\left(t\right)-y_{0}\left(t\right)\right]^{2}}}$,
in addition to $\overline{y^{2}}$ calculated earlier, we also need
to compute $\overline{y\left(t\right)}$. Introducing $\chi=\overline{\left\langle F\exp\left(i\phi\right)\right\rangle }$, from
eq.(\ref{eq:KinEqND}) we have
\[
\chi=e^{i\left(\phi_{0}+t\right)-Dt}
\]
Since $\overline{v}_{y}=\Re\chi$, for $\overline{y}\left(t\right)$
we find
\[
\overline{y}=\frac{e^{-Dt}\left[\sin\left(\phi_{0}+t\right)-D\cos\left(\phi_{0}+t\right)\right]}{D^{2}+1}+\frac{D\cos\phi_{0}-\sin\phi_{0}}{D^{2}+1}
\]
Shown in Fig.\ref{fig:Major-characteristics-of} are the plots of
$\overline{y},y_{0},\sqrt{\overline{y^{2}}}$, and the rms deviation,
$\sigma$, introduced above. It is seen that the acceptability of
the straight-line approximation, as expected, depends on $D$. For
$D\ll1$, it is applicable to relatively broad gaps, $d\lesssim r_{\text{g}0}$,
while for $D\sim1$ we have to impose a stronger condition $d\ll r_{g0}$
(where $r_{\text{g0}}$ is measured by $B_{0})$. Qualitatively, these
findings can be combined in the requirement of a small angular deflection
of particles traversing the gap $d,$ as $\Delta\vartheta\sim \sqrt{ld}/r_{\text{g}}\ll1$,
Sec.\ref{sec:Double-Shock-Acceleration-Model}. The angular deflection
can also be represented as $\Delta\vartheta\sim\sqrt{\left(d/r_{g0}\right)D\left(\delta B/B_0\right)}$. For $D\ll B_0/\delta B\ll 1$, the  $d\ll r_{\rm g0} $ constraint for the rectilinear propagation can be somewhat relaxed, 
which is consistent with our findings shown in Fig.\ref{fig:Major-characteristics-of}.
According to these results, the applicability of this approximation
can be extended by replacing the rectilinear trajectories with arcs.

A relatively mild straight-line
($\Delta\vartheta\ll1)$ propagation constraint on $r_{g}$, namely
$r_{g}\gg\sqrt{ld}$, might be violated if the shock compression and a
turbulent flow between the shocks strongly amplify the magnetic
field. At the same time, the spatial structure of such flows is often
intermittent, leading to a reduced number of effective scatterings of a particle
traversing the intershock gap, $d$, which was assumed to be $\sim d/l$
in the above estimates. For example, the turbulent field driven by
the CR current, $J_{\text{CR}},$ has been shown in {\cite{Bell05}}
to form a 3D cellular structure with relatively thin walls around
magnetic cavities of a size $a\gg l_{\text{w}}$, where $l_{\text{w}}$
is the thickness of the walls between the cavities (see also {\cite{Reville2008}}). Each cavity forms
when plasma expands from its center, being driven by the Ampere force,
$\propto J_{\text{CR}}B$. As no significant field line stretching
by turbulent eddies is observed, one may assume that the wall field
is enhanced by a factor $\sim a/l_{\text{w}}$, which reduces the local
gyroradius by the same factor. However, the field concentration in
the walls brings the number of effective scatterings between the shocks down
to $\sim d/a$. In this case, the above straight-line propagation
requirement is modified to $r_{\text{g}}\gg\sqrt{ad}$, where $r_{g}$
is still measured relative to a field close to the background field in the downstream
frame.

\section{Lorentz-Fermi Additive Groups}\label{sec:Appendix-B}

\label{sec:Lorentz-Acceleration-Transformat}

Here we approach the iterated map introduced in Sec.\ref{sec:IteratedMap}
from the additive group perspective. This approach facilitates extensions
of the map to cases of different field directions upstream of the
two shocks, different shock speeds, variable angles between the shock
surfaces, and other possible changes of shock parameters. The extensions
can be integrated into the map by inserting additional elements of
required groups. For example, if magnetic fields upstream of two shocks
make an angle $\vartheta\neq0$, we will rotate the coordinate system
around the shock normal by $\vartheta$ \emph{after} rotating it by
$\pi$ around the field direction of the shock just visited by the
particle. Then, after a particle visits the opposite shock, we rotate the system
around the field direction by $\pi$ (or $-\pi$), again. Then the
rotation by $\vartheta$ (not $-\vartheta$, because the frame is
flipped with respect to the shock plane) around the shock normal will
bring the coordinate system to the initial position for the map entering
the next cycle. As these additional elements of particle interaction
with the shocks can be straightforwardly included, we consider
only the basic field and shock geometry shown in Fig.\ref{fig:Particles-bouncing-between}.

During the excursion upstream of a shock, the particle momentum component
along the field remains constant, so we need to map only $p_{x}$
and $p_{y}$. A perpendicular momentum, $p_{\perp}=\sqrt{p_{x}^{2}+p_{y}^{2}}$,
will exceed $p_{\parallel}$ after several iterations depending
on the shock gamma factor, $\left(1-u^{2}\right)^{-1/2}$. To better
understand the group transformation corresponding to the mapping given
in eq.(\ref{eq:MapSeries}) and allow for the above-mentioned generalizations,
we will characterize the particle state between the shocks by a vector
$\boldsymbol{p}=(p_{y},\gamma,p_{x})$ in an extended (Minkowski)
phase space with a signature $(-1,1,-1)$. We have omitted the ``hats''
in this notation as it is limited to this Appendix and does not interfere
with the main text. In this space, the length square $\gamma^{2}-p_{x}^{2}-p_{y}^{2}=1+p_{\parallel}^{2}=const$
is automatically conserved under a series of transformations that
constitute a particle excursion upstream and its return to downstream
space. The first transformation is Lorentz's hyperbolic rotation:
\[
L_{\psi}=\left(\begin{array}{ccc}
\cosh\psi\; & \sinh\psi\; & 0\\
\sinh\psi\; & \cosh\psi\; & 0\\
0\; & 0\; & 1\;
\end{array}\right)
\]
where $\tanh\psi=u$ - is the shock velocity. The second transformation
performs particle momentum rotation upstream by a phase $\sigma=$$2\tau_{1}$,
where $\tau=\tau_{1}$ is the root of eq.(\ref{eq:MapTranscen}):
\[
R_{\sigma}=\left(\begin{array}{ccc}
\cos\sigma & \;0 & \;\sin\sigma\\
0 & \;1 & 0\\
-\sin\sigma & \;0 & \;\cos\sigma
\end{array}\right)
\]
Each of these two transformations forms a one-parameter additive group.
Indeed, they contain an identity transformation, represented by a
unit matrix at $\sigma=\psi=0$. The following additive group properties
are also satisfied: $L_{\psi+\chi}=L_{\psi}L_{\chi}$ and $R_{\sigma+\eta}=R_{\sigma}R_{\eta}$.
Another useful property of these groups is that the inverse matrices
are given by changing the sign of the group parameter, $\psi\to-\psi$,
$\sigma\to-\sigma$. An essential difference between these groups
is that while $L_{\psi}$ is completely determined by the shock velocity,
the group $R_{\tau}$ depends on the current particle momentum through
parameter $\tau$, which is to be obtained from the transcendental
equation given by eqs.(\ref{eq:MapTranscen}). The following three steps give the main element of
particle interaction with one shock: Lorentz transform from downstream to upstream, particle rotation
around its Larmor orbit center upstream, and another Lorentz transform
back to the downstream frame. These are given by the matrix
\begin{equation}
T_{\tau_{1}}=L_{-\psi}R_{2\tau_{1}}L_{\psi}.\label{eq:ftR}
\end{equation}
After this transformation is applied to a particle, it begins heading
to the opposite shock. Then, after applying a rotation matrix $R_{\pi}$
we may repeat the series of transformations in eq.(\ref{eq:ftR})
using parameter $\tau_{2}$ instead of $\tau_{1},$ and a different
Lorentz's angle $\psi$, if the second shock has a different speed.
The advantage of this approach is that one can easily replace the
rotation $R_{\pi}$ by a different angle, should the shocks not be
parallel to each other. This situation occurs if they are corrugated
or if the colliding flows are not strictly antiparallel. Oblique flow
collision geometry is inevitable in the colliding stellar winds, with
a possible exception for a small vicinity of the flow stagnation region.
Based on what we have discussed above, we write the full transformation
starting from the moment before entering one of the shocks and propagating
it until the particle returns to a similar position after its excursion
upstream of both shocks. The full transformation can be written as
follows
\[
T_{\tau_{2}}T_{\tau_{1}}=R_{\pi}L_{-\psi}R_{2\tau_{2}}L_{\psi}R_{\pi}L_{-\psi}R_{2\tau_{1}}L_{\psi}
\]
This construction of the iterated map is used in Sec.\ref{sec:Properties-of-Map}
(see also Fig.\ref{fig:Particles-bouncing-between}) and is valid
for the case of parallel shock planes and magnetic fields upstream.
The fixed point of the map is characterized by the condition $\tau_{1}=\tau_{2}$
with equal incidence angles. One-shock map $T_{\tau}=R_{\pi}L_{-\psi}R_{2\tau}L_{\psi}$
can be represented in an instructive form by properly commuting the
$R_{2\tau}$ and $L_{\psi}$ matrices and using the identity $L_{-\psi}L_{\psi}=1$:
\begin{align}
T_{\tau}=\left(\begin{array}{ccc}
-\cos2\tau & 0 & -\sin2\tau\\
0 & 1 & 0\\
\sin2\tau & 0 & -\cos2\tau
\end{array}\right)+\frac{1}{2}\sin^{2}\tau\;\sinh^{2}\frac{\psi}{2}\times\nonumber \\
\left(\begin{array}{ccc}
-1 & -\coth\psi & \frac{1}{2}\cot\tau\;\text{sech}^{2}\frac{\psi}{2}\\
\coth\psi & 1 & -\cot\tau\;\text{csch}\psi\\
-\frac{1}{2}\cot\tau\;\text{sech}^{2}\frac{\psi}{2} & -\cot\tau\;\text{csch}\psi & 0
\end{array}\right)\nonumber
\end{align}

The result is then automatically split into two parts: the first one
includes only the particle rotation upstream and the rotation of the
coordinate system by the angle $\pi$, since the Lorentz's transformations
cancel out. The second part comes from the product of Lorentz transform
and particle rotation upstream, which produces the energy increase.
While it is obvious but still worth remarking that the determinant
$\left|T_{\tau}\right|=1$, thus conforming to the space phase volume
conservation (Liouville's theorem) upon mapping. As the independent
part of this map applies to the particle momentum components $p_{x},p_{y}$,
this map is an area preserving map. As the map stretches the momentum
$p$ it must contract the angular distribution of mapped particles,
as seen in Fig.\ref{fig:AngleEnergyMaps}, for example.

\bibliographystyle{apsrev4-1}
\bibliography{FinalArxivVer}

\end{document}